\documentclass[11pt,showpacs,superscriptaddress,amsmath,amssymb,nofootinbib]{revtex4}

\usepackage{graphicx}
\usepackage{dcolumn}
\usepackage{bm}
\usepackage{amssymb}
\usepackage{amsmath}
\usepackage{epsfig}
\usepackage{color}
\usepackage{slashed}
\usepackage{float}
\usepackage{longtable}
\usepackage{braket}
\usepackage{ulem}
\def\w{\omega}
\def\e{\eta}
\def\r{\rho}
\def\f{\phi}
\def\p{\pi}
\def\o{\overline}
\def\d{\delta}
\newcolumntype{I}{!{\vrule width 1.3pt}}
\begin{document}

\title{Updating $B \to PP,VP$ decays in the framework of flavor symmetry}

\author{Hai-Yang Cheng}
\email{phcheng@phys.sinica.edu.tw}
\affiliation{Institute of Physics, Academia Sinica, Taipei, Taiwan 11529, Republic of China}
\author{Cheng-Wei Chiang}
\email{chengwei@ncu.edu.tw}
\affiliation{Institute of Physics, Academia Sinica, Taipei, Taiwan 11529, Republic of China}
\affiliation{Department of Physics and Center for Mathematics and Theoretical Physics,
National Central University, Chungli, Taiwan 32001, Republic of China}
\affiliation{Physics Division, National Center for Theoretical Sciences, Hsinchu, Taiwan 30013, Republic of China}
\author{An-Li Kuo}
\email{101222028@cc.ncu.edu.tw}
\affiliation{Department of Physics and Center for Mathematics and Theoretical Physics,
National Central University, Chungli, Taiwan 32001, Republic of China}

\begin{abstract}
Current data of charmless $B$ meson decays to two pseudoscalar mesons ($PP$) and one vector and one pseudoscalar mesons ($VP$) are analyzed within the framework of flavor SU(3) symmetry, a working principle that we have tested by allowing symmetry breaking factors in the decay amplitudes and found to be a good approximate symmetry.  In the $PP$ sector, the color-suppressed tree amplitude is found to be larger than previously known and has a strong phase of $\sim -70^\circ$ relative to the color-favored tree amplitude.  We have extracted for the first time the $W$-exchange and penguin-annihilation amplitudes.  The former has a size of about the QCD-penguin amplitude and a phase opposite to that of the color-favored tree amplitude, while the latter is suppressed in magnitude but gives the dominant contribution to the $B_s^0\to \pi^+\pi^-$ and $\pi^0\pi^0$ decays.
In the $VP$ sector, one striking feature is that the color-suppressed tree amplitude with the spectator quark ending up in the vector meson has a large size and a strong phase of $\sim -90^\circ$ relative to the color-favored tree amplitudes.  The associated electroweak penguin amplitude also has a similar strong phase and a magnitude comparable to the corresponding QCD penguin amplitude.  This leads to a large branching fraction of order $10^{-6}$ for $B_s^0\to \phi\pi^0$.
In contrast, the color-suppressed tree, QCD penguin, and electroweak penguin amplitudes with the spectator quark ending up in the pseudoscalar meson have magnitudes more consistent with na{\"i}ve expectations.  Besides, current data are not sufficiently precise for us to fix the $W$-exchange amplitudes.
For both the $PP$ and $VP$ sectors, predictions of all the decay modes are made based upon our preferred fit results and compared with data and those made by perturbative approaches.  We have identified a few observables to be determined experimentally in order to discriminate among theory calculations.
\end{abstract}


\maketitle
\newpage

\section{Introduction \label{sec:intro}}

Thanks to experimental efforts in the past decade or so, branching fractions and CP asymmetries of most charmless $B_{u,d}$ meson decays to two pseudoscalar mesons ($PP$) and one vector and one pseudoscalar mesons ($VP$) had been measured.  Those of a few $B_s$ decays were also observed.  Such information has provided an ideal realm for us to test our theoretical understanding of heavy quark systems as well as to put constraints on new physics interactions.  Before the LHCb resumes its flavor physics program and the super B factory starts its operations, both running at higher sensitivities and statistics, it is timely to examine current data on these decay modes, check their consistency, and make predictions for observables of yet observed ones, particularly the $B_s$ decays.

Based on effective field theories, there are three major QCD-inspired approaches to hadronic $B$ decays; namely, the QCD factorization (QCDF)~\cite{QCDF}, perturbative QCD (pQCD)~\cite{pQCD}, and soft-collinear effective theory (SCET)~\cite{SCET}.  They differ in the treatment of dynamical degrees of freedom at different mass scales.  Nevertheless, factorization for hadronic matrix elements of tree-level processes is proved at the leading order in $\Lambda_{QCD}/m_b$, where $\Lambda_{QCD}$ and $m_b$ denote respectively the typical hadronic scale and the $b$ quark mass.

In contrast to the perturbative analysis, the flavor diagram approach~\cite{SU3F} is non-perturbative in nature.  It makes use of flavor SU(3) symmetry to relate decay diagrams, both sizes and associated strong phases, of the same topology but differing in the light quarks.  One advantage of this approach is to extract the decay matrix elements directly from data without reference to any specific model.  In particular, the theory parameters extracted from data in this formalism encompass effects of strong interactions to all orders, including long-distance rescattering as well.  In the past, we have thereby gained valuable knowledge about strong dynamics in various decay diagrams.  For example, the color-suppressed diagram is known to be larger than na\"{i}vely expected and has a sizeable strong phase that cannot be calculated from first principles.  Though a challenge for theorists, this has taught us that our current understanding of quantum chromodynamics (QCD) at the low energies is insufficient.

Quite a few analyses of rare hadronic $B$ decays in the flavor diagram approach~\cite{GlobalFits,cwPP,cwPV} had been done before based on the available data then.  In this work, we want to update the analyses using the latest data.  With more and better determined data than before, we observe for the first time the need of the $W$-exchange and penguin annihilation amplitudes in the $PP$ decays.  As another example, we find one electroweak penguin amplitude in the $VP$ decays larger than na{\"i}ve expectations.  It is therefore worth studying what are the implications of such new findings.  More importantly, based on the theory parameters extracted from $\chi^2$ fits, we make predictions for yet measured observables and compare with those made by QCDF, pQCD, and SCET calculations.

This paper is organized as follows.  In Section~\ref{sec:formalism}, we review the flavor diagram approach employed in our analysis, listing all the flavor amplitudes considered in this work.  In Section~\ref{sec:fitting}, we describe the general procedure of $\chi^2$ fits to determine the size and strong phase of each flavor amplitude using the latest experimental data.   As no significant quantitative changes in the extracted theory parameters are found when symmetry breaking factors are introduced, we choose to present only the fit results under exact flavor SU(3) symmetry.  Afterwards, we divide our analyses into two parts: Section~\ref{sec:PP} for the $PP$ sector and Section~\ref{sec:VP} for the $VP$ sector.  For each sector, we first present experimental data and flavor amplitude decomposition for each mode, followed by the results of theory parameters extracted from $\chi^2$ fits to $B_{u,d}$ decays in various schemes differing in whether certain modes and/or flavor diagrams are included or not.  Measured observables in the $B_s$ decays are purposely left out from the fits to test the flavor symmetry.  We discuss implications of these results and consider different fit schemes when necessary.  Finally, we make predictions for the branching fractions and CP asymmetries of all the decay modes based on the preferred fit results.  A comparison between our predictions and others' can be found at the end of each section.  In Section~\ref{sec:PPQCDFsec}, we compute the effective Wilson coefficients $a_1$ and $a_2$ for a few representative modes and compare them with values derived from perturbation approaches.  Conclusions of our work are given in Section~\ref{sec:conclusions}.

\section{Flavor Diagram Approach \label{sec:formalism}}

Transition amplitudes for heavy meson decays can be categorized according to their flavor flow topologies.  Among these flavor diagrams, seven types had been identified to play indispensable roles in explaining the data.  Leaving out the Cabibbo-Kobayashi-Maskawa (CKM) factors, they are:
\begin{itemize}
\item $T$, denoting the color-favored tree diagram with external $W$ emission;
\item $C$, denoting the color-suppressed tree diagram with internal $W$ emission;
\item $E$, denoting the $W$-exchange diagram;
\item $P$, denoting the QCD penguin diagram;
\item $S$, denoting the flavor-singlet QCD penguin diagram;
\item $P_{EW}$, denoting the electroweak (EW) penguin diagram;
\item $P\!A$, denoting the penguin annihilation diagram.
\end{itemize}
$T$ and $C$ are expected to be the most dominant amplitudes, with $C$ being na\"{i}vely smaller than $T$ by a color factor of 3.  $E$ is suppressed by helicity and/or hadronic form factors.  The rest four types of amplitudes are suppressed by loop factors.  Compared to the first five types of diagrams, the EW penguin diagram is one order higher in weak interactions and thus even smaller in strength.  As we will see, however, current data show a less clear hierarchy as mentioned above.  This is a hint of possibly non-perturbative strong dynamics at play.  The above seven flavor diagrams are sufficient to explain the observed data for the $PP$ modes.  In the case of the $VP$ modes, both the $W$-exchange and the penguin annihilation diagrams are not called for by data at the current precision level.  Otherwise, the number of flavor diagrams is doubled.  This is because one has to distinguish cases where the spectator quark in the $B$ meson ends up in the vector or pseudoscalar meson in the final state.  The corresponding flavor diagram symbols are added with a subscript $V$ or $P$, respectively.  These two sets of amplitudes are different {\it a priori}.  Yet they can be related to each other under the assumption of factorization.  Each amplitude mentioned above can be factored as its modulus multiplied by an associated strong phase.  Moreover, we take the convention of fixing $T$ (in the case of $PP$ decays) and $T_P$ (in the case of $VP$ decays) to be real, and all the other strong phases, denoted by $\d_X$ for amplitude $X$, are relative to these amplitudes.  For completeness, we will also include in the following flavor amplitude decomposition the color-suppressed EW penguin diagram $P^C_{EW}$ that is both loop-suppressed and sub-leading in weak interactions, thereby not taken into account in our numerical analyses.

We fix the phase convention of the iso-doublet anti-quarks in such a way that $({\o d},-{\o u})^T$ transforms exactly the same as $(u,d)^T$~\cite{Halzen:1984mc} for the convenience of isospin symmetry analysis.
As a result, the quark contents for light pseudoscalar mesons are $\pi^+=u\o d$, $\pi^0=(d\o d-u\o u)/\sqrt2$, $\pi^-=-d\o u$, $K^+=u\o s$, $K^0=d\o s$, $\o K^0=s\o d$, $K^-=-s\o u$ and those for light vector mesons are $\rho^+=u\o d$, $\rho^0=(d\o d-u\o u)/\sqrt2$, $\rho^-=-d\o u$, $K^{*+}=u\o s$, $K^{*0}=d\o s$, $\o K^{*0}=s\o d$, $K^{*-}=-s\o u$, $\omega=(u\o u+d\o d)/\sqrt2$ and $\phi=s\o s$.  The physical $\eta$ and $\eta'$ mesons are mixtures of $\eta_q=\frac{1}{\sqrt2}(u\o u+d\o d)$ and $\eta_s=s\o s$ in the following way~\cite{mixingangle}:
\begin{equation}
\begin{pmatrix}
\eta \\
\eta'
\end{pmatrix}
=
\begin{pmatrix}
\cos\phi  & -\sin\phi \\
\sin\phi  &  \cos\phi \\
\end{pmatrix}
\begin{pmatrix}
\eta_q \\
\eta_s
\end{pmatrix} ~,
\end{equation}
where the mixing angle $\phi$ is fixed at $46^\circ$~\cite{phi} for subsequent analyses.

In physical processes, the above-mentioned flavor amplitudes always appear in certain combinations, multiplied by appropriate Cabibbo-Kobayashi-Maskawa (CKM) factors.  Therefore, we introduce small letters to denote these combinations:
\begin{align}
t &=Y^u_{db}T-(Y^u_{db}+Y^c_{db})P^C_{EW} ~,       &    t'&= Y^u_{sb}\xi_tT-(Y^u_{sb}+Y^c_{sb})P^C_{EW} ~,
\nonumber \\
c &=Y^u_{db}C-(Y^u_{db}+Y^c_{db})P_{EW} ~,         &    c'&= Y^u_{sb}\xi_cC-(Y^u_{sb}+Y^c_{sb})P_{EW} ~,
\nonumber \\
e &=Y^u_{db}E ~,                                                       &    e'&=Y^u_{sb}E ~,
\nonumber \\
p &=-(Y^u_{db}+Y^c_{db})(P-\frac{1}{3}P^C_{EW}) ~, &    p'&=-(Y^u_{sb}+Y^c_{sb})(\xi_pP-\frac{1}{3}P^C_{EW}) ~,
\\
s &=-(Y^u_{db}+Y^c_{db})(S-\frac{1}{3}P_{EW}) ~,   &    s'&=-(Y^u_{sb}+Y^c_{sb})(\xi_sS-\frac{1}{3}P_{EW}) ~,
\nonumber \\
pa &=-(Y^u_{db}+Y^c_{db})P\!A ~,       &    pa'&=-(Y^u_{sb}+Y^c_{sb})P\!A ~,
\nonumber
\end{align}
where unprimed and primed amplitudes represent strangeness-conserving ($\Delta S=0$) and strangeness-changing ($|\Delta S|=1$) transitions, respectively, and $Y^{q'}_{qb}\equiv V_{q'q}V^*_{q'b}$ with $q,q' = u,d,s$ and $V_{qq'}$ being a CKM matrix element.  In the case of penguin amplitudes, we have utilized the unitarity relation to integrate out the top quark.  Moreover, the factors $\xi_{t,c,p,s}$ are introduced as SU(3) breaking factors for the corresponding flavor diagrams when going from $\Delta S=0$ transitions to $|\Delta S|=1$ transitions.  They are unity in the limit of flavor $SU(3)$ symmetry.  One working assumption here is that the strong phase of each flavor diagram is identical for $\Delta S=0$ and $|\Delta S|=1$ ones.

Through a Fierz transformation, the EW penguin operators contributing to $P_{EW}$ and $P^C_{EW}$ can be related to the tree operators responsible for $T$ and $C$~\cite{EWP}, leading to the relations
\begin{equation}
P_{EW}=-\delta_{EW}|T|e^{i\delta_{P_{EW}}} ~\mbox{ and }~ P^C_{EW}=-\delta_{EW}|C|e^{i\delta_{P^C_{EW}}} ~,
\end{equation}
where, in terms of the Wilson coefficients $C_i$~\cite{Buchalla:1995vs},
\begin{align}
\delta_{EW} \simeq \frac32 \frac{C_9 + C_{10}}{C_1 + C_2} \simeq 0.0135 \pm 0.0012
\end{align}
from perturbative calculations.  This is smaller than what we find from data.

We will employ the Wolfenstein parameterization for the CKM matrix elements.  Since the Wolfenstein parameters $A$, $\lambda$, $\o\rho \equiv \rho(1-\frac{\lambda^2}{2})$ and $\o\eta \equiv \eta(1-\frac{\lambda^2}{2})$ have been determined to a high precision by other processes, we simply adopt their central values given by the CKMfitter Group~\cite{Charles:2004jd}:
\begin{align}
&
A =0.813^{+0.015}_{-0.027}
~,~~
\lambda =0.22551^{+0.00068}_{-0.00035}
~,~~
\o\rho =0.1489^{+0.0158}_{-0.0084}
~,~~
\o\eta =0.342^{+0.013}_{-0.011} ~.
\end{align}
We also take the central values of the $B$ meson lifetimes $\tau_{B^+}=(1.641\pm0.008)$ ps, $\tau_{B^0}=(1.519\pm0.007)$ ps and $\tau_{B_s}=(1.497\pm0.015)$ ps~\cite{PDG}.

\section{General Aspects of Data Fitting \label{sec:fitting}}

For a two-body $B$ meson decay process, the decay width is given by
\begin{equation}
\Gamma(B\to M_1M_2)=\frac{p}{8\pi m_B^2}|{\cal M}|^2 ~,
\label{decaywidth}
\end{equation}
where $m_B$ is the $B$ meson mass, $p$ denotes the magnitude of the 3-momentum of either meson in the final state, $M_{1,2}$ can be either a pseudoscalar or a vector meson, and $\cal M$ represents the corresponding decay amplitude.  The branching fraction of each mode is obtained by multiplying the CP-averaged partial width, $\o\Gamma \equiv \left[\Gamma(B\to M_1M_2) + (\o B\to \o M_1 \o M_2)\right]/2$, by the $B$ meson lifetime.  The direct CP asymmetry is defined as
\begin{align}
A_{CP}(B\to M_1M_2) =
\frac{\Delta\Gamma(B\to M_1M_2)}{\o\Gamma(B\to M_1M_2)} ~,
\end{align}
where $\Delta\Gamma(B\to M_1M_2) \equiv \Gamma(\o B\to \o M_1 \o M_2) - \Gamma(B\to M_1M_2)$.  In the case where a neutral $B$ meson and its charge conjugate can decay into the same final state $f_{CP}$, the associated time-dependent CP asymmetry is defined as
\begin{equation}
A_{CP}(t) =
\frac{\Gamma(\o B^0\to f_{CP})-\Gamma(B^0\to f_{CP})}{\Gamma(\o B^0\to f_{CP})+\Gamma(B^0\to f_{CP})}
={\cal S}\sin(\Delta m_B\, t) + {\cal A}\cos(\Delta m_B\, t)
\label{eq:tACP}
\end{equation}
where $\Delta m_B$ is the difference between the two mass eigenvalues of the neutral $B$ mesons, $\cal S$ is the mixing-induced CP asymmetry, and $\cal A$ is the direct CP asymmetry.  These time-dependent CP asymmetries are calculated to be
\begin{equation}
{\cal A} =\frac{|\lambda_f|^2-1}{1+|\lambda_f|^2}
~\mbox{ and }~
{\cal S}=\frac{2~\text{Im}[\lambda_f]}{1+|\lambda_f|^2} ~,
\label{AandS}
\end{equation}
where
\begin{align}
\lambda_f=\frac{q}{p}\frac{\o A_f}{A_f}
~\mbox{ and }~
\frac{q}{p}=\frac{V_{tb}^*V_{td}}{V_{tb}V_{td}^*}
~\left( \mbox{or } \frac{V_{tb}^*V_{ts}}{V_{tb}V_{ts}^*} \right)
\end{align}
for $B^0$ (or $B_s$) meson decays, $A_f$ denotes the $B^0 \to f_{CP}$ decay amplitude and $\o A_f$ the conjugate amplitude.

In our approach, the branching fractions and CP asymmetries of decay modes become functions of the moduli and strong phases of the flavor amplitudes.  We extract these theory parameters through a $\chi^2$ fit to data.
Uncertainties of the experimental data, including the scale factor when applicable, are used in the fits.  After finding the parameters that render the minimal $\chi^2$ value, $\chi^2_{\rm min}$, we take them as the central values and scan for their 1-sigma ranges.  We have done full standard deviation scans and observed that the correlations among the parameters are sufficiently small and would lead to tiny differences in predictions.  Therefore, for simplicity and convenience in presentation, our results below assume no correlations in the theory parameters.

As to the experimental data, we quote mostly the world-averaged results given by the Heavy Flavor Averaging Group (HFAG)~\cite{HFAG} and new data from the LHCb Collaboration \cite{KpiKK,BsKKpipi,phiKphipi,BsVPexp}, the Belle Collaboration \cite{omegaK} and the recent ICHEP updates \cite{pi0pi0exp,etaKexp}.  When there is a large discrepancy among data of different experimental groups, we do the weighted average by ourselves and include a scale factor in the standard deviation.

In our fits, we only make use of the observables in the decays of $B^+$ and $B^0$ mesons, as data of the $B_s$ decays are comparatively scarce.  Moreover, we generally divide our fits into two categories: one being restricted to the decay modes involving no flavor-singlet diagrams (Schemes A and B in the $PP$ sector and Scheme A in the $VP$ sector), and the other being for all the decay modes (Schemes C and D in the $PP$ sector and Schemes B and C  in the $VP$ sector).  The former restricted fits avoids the uncertainty in the $\eta$-$\eta'$ mixing, and serves as a guide to looking for a reasonable solution in the latter global fits.

\section{The $B\to PP$ Sector \label{sec:PP}}

\begingroup
\begin{table}[t!]
\begin{tabular}{clccc}
\hline\hline
\multicolumn{2}{c}{Mode} & Flavor Amplitude & ~~~$BF$~~~ & ~~~$A_{CP}$~~~ \\
\hline
$B^+ \to$
& $\pi^+\pi^0$     & $-\frac{1}{\sqrt2}(t+c)$                                                                                & $5.48^{+0.35}_{-0.34}$         &$0.026\pm0.039$\\
&$K^+\o K^0$    & $p$                                                                                                              & $1.19\pm0.18$ (1.02)            &$-0.086\pm0.100$ \cite{KpiKK}\\
&$\e\pi^+$          & $\frac{c_\phi}{\sqrt2}[t+c+2p+(2-\sqrt2t_\phi)s]$                                        & $4.02\pm0.27$                      &$-0.14\pm0.05$ (1.42)\\
&$\e'\pi^+$         & $\frac{s_\phi}{\sqrt2}[t+c+2p+(2+\frac{\sqrt2}{t_\phi})s]$                             & $2.7^{+0.5*}_{-0.4}$ (1.36)    &$0.06\pm0.15^*$\\
\hline
$B^0 \to$
&$K^+K^-$         & $-(e+2pa)$                                                                                                   & $0.12\pm0.05$                   &-\\
&$K^0\o K^0$    & $p+2pa$                                                                                                      & $1.21\pm0.16$                      &$0.06\pm0.26$ (1.38)\\
     &                        &                                                                                                                 &                                               &$-1.08\pm0.49$\\
&$\pi^+\pi^-$      & $-(t+p+e+2pa)$                                                                                            & $5.10\pm0.19$                      &$0.31\pm0.05$ \cite{BsKKpipi}\\
     &                        &                                                                                                                                 &                                               &$-0.66\pm0.06$ \cite{BsKKpipi}\\
&$\pi^0\pi^0$     & $\frac{1}{\sqrt2}(-c+p+e+2pa)$                                                                     &$1.17\pm0.13$ (3.18) \cite{pi0pi0exp}  &$0.03\pm0.17$ (1.94) \cite{pi0pi0exp}\\
&$\e\pi^0$         & $\frac{c_\phi}{2}[2p+(2-\sqrt2t_\phi)s-2e]$                                                     & $<1.5$                                &-\\
&$\e'\pi^0$        & $s_\phi[p+(1+\frac{1}{\sqrt2t_\phi})s-e]$                                        & $1.2\pm0.4$ (1.46)               &-\\
&$\e\e$             & $\frac{c^2_\phi}{\sqrt2}[c+p+(2-\sqrt2t_\phi)s+e+\frac{2}{c_\phi^2}pa]$              & $<1.0$                                &-\\
&$\e'\e$            & $\frac{c_\phi s_\phi}{2}[2c+2p+(4-\sqrt2t_\phi+\frac{\sqrt2}{t_\phi})s+2e]$    & $<1.2$                                &-\\
&$\e'\e'$           & $\frac{s^2_\phi}{\sqrt2}[c+p+(2+\frac{\sqrt2}{t_\phi})s+e+\frac{2}{s_\phi^2}pa]$& $<1.7$           &-\\
        \hline
$B_s \to$
&$\pi^+K^-$     & $-(t+p)$                                                                                                           & $5.4\pm0.6^*$                      &$0.26\pm0.04^*$\\
&$\pi^0\o K^0$& $\frac{1}{\sqrt2}(-c+p)$                                                                                    & -                                            &-\\
&$\e\o K^0$     & $\frac{c_\phi}{\sqrt2}[c+(1-\sqrt2t_\phi)p+(2-\sqrt2t_\phi)s]$                          & -                                            &-\\
&$\e'\o K^0$    & $\frac{s_\phi}{\sqrt2}[c+(1+\frac{\sqrt2}{t_\phi})p+(2+\frac{\sqrt2}{t_\phi})s]$ & -                                            &-\\
\hline\hline
\end{tabular}\\
\caption{Flavor amplitude decomposition, branching fractions ($BF$) in units of $10^{-6}$ and CP asymmetries ($A_{CP}$) for strangeness-conserving $B\to PP$ decays.  When there are more than one line for a decay mode, the CP asymmetry in the upper line is $\cal A$ and that in the lower line is $\cal S$, both defined in Eq.~(\ref{eq:tACP}). The short-hand notations $s_\phi$, $c_\phi$ and $t_\phi$ are used to denote $\sin\phi$, $\cos\phi$, and $\tan\phi$, respectively.  When there is a significant discrepancy among data from different experimental groups, the error for that entry is enlarged by the corresponding scale factor given in parentheses.  We use an asterisk to label each observable not taken into account in our analysis, with reasons given in the text.}
\label{PPamplitude0}
\end{table}
\endgroup

\begingroup
\begin{table}[t!]
\begin{tabular}{lcccc}
\hline\hline
\multicolumn{2}{c}{Mode}  & Flavor Amplitude & ~~~$BF$~~~ & ~~~$A_{CP}$~~~   \\
\hline
$B^+ \to$
&$K^0\pi^+$              & $p'$                                                                                                                                                       &$23.79\pm0.75$                     &$-0.017\pm0.016$ \cite{KpiKK}\\
&$K^+\pi^0$              & $-\frac{1}{\sqrt2}(p'+t'+c')$                                                                                                                    &$12.94^{+0.52}_{-0.51} $        &$0.040\pm0.021$\\
&$\e K^+$                 & $\frac{c_\phi}{\sqrt2}[t'+c'+(1-\sqrt2t_\phi)p'+(2-\sqrt2t_\phi)s']$                                                          & $2.36^{+0.22}_{-0.21}$ (1.18)&$-0.37\pm0.08$\\
&$\e'K^+$                 & $\frac{s_\phi}{\sqrt2}[t'+c'+(\frac{\sqrt2}{t_\phi}+1)p'+(2+\frac{\sqrt2}{t_\phi})s']$                                  & $71.1\pm2.6$                         &$0.013\pm0.017$\\
\hline
$B^0 \to$
&$K^+\pi^-$              & $-(p'+t')$                                                                                                                                                 & $19.57^{+0.53}_{-0.52}$       &$-0.082\pm0.006$\\
&$K^0\pi^0$             & $\frac{1}{\sqrt2}(p'-c')$                                                                                                                            & $9.93\pm0.49$                      &$-0.01\pm0.10$ (1.38)\\
      &                               &                                                                                                                                                                &                                               &$0.57\pm0.17$\\
&$\e K^0$                 & $\frac{c_\phi}{\sqrt2}[c'+(1-\sqrt2t_\phi)p'+(2-\sqrt2t_\phi)s']$                                                               & $1.23^{+0.27}_{-0.24}$         &-\\
&$\e'K^0$                 & $\frac{s_\phi}{\sqrt2}[c'+(\frac{\sqrt2}{t_\phi}+1)p'+(2+\frac{\sqrt2}{t_\phi})s']$                                       & $66.1\pm3.1$ (1.32)              &$0.05\pm0.04$ \cite{etaKexp}\\
      &                               &                                                                                                                                                                &                                              &$0.63\pm0.06$ \cite{etaKexp}\\
          \hline
$B_s \to$
&$K^+K^-$                 & $-(p'+t'+e'+2pa')$                                                                                                                                       & $24.5\pm1.8^*$                   &$-0.14\pm0.11^*$ \cite{BsKKpipi}\\
     &                                &                                                                                                                                                                &                                             &$0.30\pm0.13^*$ \cite{BsKKpipi}\\
&$K^0\o K^0$            & $p'+2pa'$                                                                                                                                                  & $<66^*$                               &-\\
&$\pi^+\pi^-$              & $-(e'+2pa')$                                                                                                                                              & $0.73\pm0.14^*$ (1.30)      &-\\
&$\pi^0\pi^0$              & $\frac{1}{\sqrt2}(e'+2pa')$                                                                                                                        & -                                          &-\\
&$\e\pi^0$                   &$-\frac{c_\phi}{2}[-\sqrt2t_\phi c'+2e']$                                                                                                         & -                                         &-\\
&$\e'\pi^0$                  &$-\frac{s_\phi}{2}[\frac{\sqrt2}{t_\phi}c'+2e']$                                                                                               & -                                         &-\\
&$\e\e$                       & $s_\phi c_\phi[-c'+\sqrt2t_\phi p'+(\sqrt2t_\phi-2)s'+\frac{e'}{\sqrt2t_\phi}+\frac{\sqrt2}{c_\phi s_\phi}pa']$&-                             &-\\
&$\e\e'$                      &$-c_\phi s_\phi[(\frac{t_\phi}{\sqrt2}-\frac{1}{\sqrt2t_\phi})c'+2p'+(\sqrt2t_\phi-\frac{\sqrt2}{t_\phi}+2)s'-e']$& -                         &-\\
&$\e'\e'$                     & $c_\phi s_\phi[c'+\frac{\sqrt2}{t_\phi}p'+(2+\frac{\sqrt2}{t_\phi})s'+\frac{t_\phi}{\sqrt2}e'+\frac{\sqrt2}{c_\phi s_\phi}pa']$ & -      &-\\
\hline\hline
\end{tabular}\\
\caption{Same as Table~\ref{PPamplitude0} but for strangeness-changing $B \to PP$ decays.}
\label{PPamplitude1}
\end{table}
\endgroup

Current experimental data on branching fractions and CP asymmetries as well as the flavor amplitude decomposition for all the $B \to PP$ decays are given in Table~\ref{PPamplitude0} and Table~\ref{PPamplitude1} for strangeness-conserving and strangeness-changing transitions, respectively.  According to our prescription in Section~\ref{sec:formalism}, there are totally 13 theory parameters to fit in this sector.  Due to the hierarchy in CKM factors, the $T$, $C$ and $E$ amplitudes are mainly determined by the $|\Delta S|=0$ transitions and the $P$, $S$ and $P_{EW}$ amplitudes by the $|\Delta S|=1$ transitions.

Four schemes of fitting are performed in our analysis.  In Schemes A and B, we do restricted fits to data without and with the $E$ and $PA$ amplitudes, respectively.  In a similar fashion, we work out global fits to data in Schemes C and D, but with the $S$ amplitude also taken into account.  Since the $B^0\to K^+K^-$ decay involves only the $E$ and $P\!A$ amplitudes, this mode is left out in Schemes A and C.  In our trial fits, we find that the observables of the $B^+\to\eta'\pi^+$ decay have large contributions to the $\chi^2$ value.  Removing them does not change the values of theory parameters much while the fit quality improves significantly.  We therefore do not include them in the fits, either.  In summary, we have 21 observables for 7 parameters in Scheme A, 22 observables for 11 parameters in Scheme B, 32 observables for 9 parameters in Scheme C, and 33 observables for 13 parameters in Scheme D.  We have tried to vary the symmetry breaking factors $\xi$'s, but observed no significant deviations from unity and not much change in fit quality.  Therefore, we choose to present only the results with exact flavor SU(3) symmetry; {\it i.e.,} the SU(3) breaking factors $\xi$'s are fixed at unity.

\subsection{Fit Results}

\begingroup
\begin{table}[t!]
\begin{tabular}{ccccc}
\hline\hline
~~Parameter~~ & ~~Scheme A~~                 & ~~Scheme B~~               & ~~Scheme C~~                     & ~~Scheme D~~ \\
\hline
$|T|$                & $0.625^{+0.013}_{-0.014}$ &$0.692^{+0.054}_{-0.085}$ &$0.627^{+0.013}_{-0.014}$  &$0.690^{+0.049}_{-0.062}$\\
$|C|$                & $0.500\pm0.049$                &$0.480^{+0.087}_{-0.084}$ & $0.607^{+0.036}_{-0.037}$ &$0.608\pm0.054$\\
$\d_C$             & $-60^{+9}_{-8}$                   & $-68\pm9$                         &$-77\pm5$                            &$-83^{+6}_{-5}$\\
$|P|$                 & $0.123\pm0.001$               & $0.124\pm0.001$              &$0.124\pm0.001$                 &$0.124\pm0.001$\\
$\d_P$              & $-24\pm2$                          & $-22^{+2}_{-4}$                 &$-24\pm2$                           &$-22^{+2}_{-3}$\\
$|P_{EW}|$       & $0.012^{+0.005}_{-0.002}$&$0.011^{+0.004}_{-0.002}$ & $0.018^{+0.006}_{-0.005}$ &$0.020\pm0.006$\\
$\d_{P_{EW}}$ & $-6^{+29}_{-42}$                & $-23^{+40}_{-39}$              &$-77^{+20}_{-11}$                &$-81^{+16}_{-9}$\\
$|E|$                & -                                           &$0.098^{+0.022}_{-0.024}$ &-                                           &$0.101^{+0.020}_{-0.022}$\\
$\delta_E$       & -                                           &$-135^{+52}_{-44}$             &-                                           &$-129^{+36}_{-32}$\\
$|P\!A|$              & -                                           &$0.011^{+0.004}_{-0.006}$ &-                                         &$0.012\pm0.004$\\
$\delta_{P\!A}$   & -                                          &$-123^{+27}_{-25}$             &-                                         &$-130^{+23}_{-21}$\\
$|S|$                & -                                           &  -                                        &$0.080\pm0.007$                 &$0.079\pm0.006$\\
$\delta_S$       & -                                           & -                                         &$-101\pm6$                          &$-98\pm6$\\
\hline
$\chi^2_{min}/dof$ & 23.41/14                        & $19.48/11$                        &$45.80/23$                            &$37.08/20$\\
Fit quality         & $5.40\%$                             & $5.30\%$                          &$0.32\%$                               &$1.14\%$\\
\hline
$\d_{EW}$       &$0.019\pm0.006$                 & $0.016\pm0.004$             &$0.029\pm0.009$                   &$0.029\pm0.009$\\
$|C/T|$            &$0.80\pm0.08$                      & $0.69\pm0.14$                  &$0.97\pm0.06$                      &$0.89\pm0.11$\\
\hline\hline
\end{tabular}\\
\caption{Fit results of theory parameters.  Only the $\pi\pi$, $\pi K$ and $KK$ decay modes are used in Schemes A and B, while Schemes C and D include all available $PP$ observables in the $B^{+,0}$ decays.  Magnitudes of the amplitudes are quoted in units of $10^4$ eV, and the strong phases in units of degree.  The branching fraction of $B^0\to K^+K^-$ is taken into account only in Schemes B and D.}
\label{PPpara}
\end{table}
\endgroup

Table~\ref{PPpara} summarizes the results of our four fits.  The amplitudes show the following pattern in size: $|T|  \agt |C| > |P|, |E| > |S| > |P_{EW}|\sim |P\!A|$.  With the inclusion of $E$ and $P\!A$ amplitudes in the restricted fits, we do not observe much change in the fit quality, as shown by our results of Schemes A and B.  However, the global fit of Scheme D is about 3 times better than that of Scheme C, indicating the importance of the $E$ and $P\!A$ amplitudes.  Their constraints come from the data of $B^0 \to K^+ K^-$, $\pi^+ \pi^-$ and $\pi^0 \pi^0$ decays.  The $E$ amplitude is seen to have a size about the same as the $P$ amplitude and a phase of $\sim -130^\circ$ relative to the $T$ amplitude.  On the other hand, the $P\!A$ amplitude has a similar phase as $E$ but is one order of magnitude smaller in size than $P$.

\begingroup
\begin{table}[t!]
\begin{tabular}{ccccc}
\hline\hline
 ~~Parameter~~ & ~~Scheme B~~ & Remove $\pi^0\pi^0$ & Remove $K^0\pi^0$\\
                           &  & $BF$ and $\cal A$ & $BF$, $\cal A$ and $\cal S$ \\
\hline
   $|T|$                      & 0.692         &0.731          & 0.684 \\
   $|C|$                      & 0.480         &0.527          &0.493\\
   $\delta_C$             & $-68$         & $-79$         & $-68$    \\
   $|P|$                      & 0.124         &0.124           & 0.123 \\
   $\delta_P$             & $-22$         & $-21$          & $-22$      \\
   $|P_{EW}|$            & 0.011         &0.014	          & 0.014 \\
   $\delta_{P_{EW}}$ & $-23$        & $-59$          & $-28$    \\
   $|E|$                      & 0.097         &0.108           & 0.096 \\
   $\delta_E$             & $-135$       &$172$          &$-130$      \\
    $|P\!A|$                   & 0.011         &0.004           & 0.012 \\
   $\delta_{PA}$         & $-123$       &$-117$         &$-121$      \\
\hline
    $\chi^2_{min}/dof$& $19.48/11$ &$16.65/9$  & $15.91/8$ \\
   Fit quality               & $5.30\%$    &$5.45\%$  & $4.37\%$\\
\hline
     $|C/T|$                 & 0.69           &0.72            &0.72    \\
     $BF(\pi^0\pi^0)$   &1.43            &2.09            &1.43      \\
     ${\cal A}(\pi^0\pi^0)$  & 0.354       &0.591           & 0.365  \\
     ${\cal S}(\pi^0\pi^0)$  & 0.791       &0.486            & 0.768    \\
     $BF(K^0\pi^0)$     & 9.55         &9.58              & 9.00      \\
     ${\cal A}(K^0\pi^0)$    &$-0.105$       &$-0.142$           & $-0.113$  \\
     ${\cal S}(K^0\pi^0)$   &0.783        &0.764            & 0.785    \\
         \hline
      \end{tabular}\\
\caption{Results of fits after taking away several observables in Scheme B.  Only the central values of theory parameters and predicted observables are shown.  Magnitudes of the amplitudes are given in units of $10^4$ eV, the strong phases in units of degree and the branching fractions in units of $10^{-6}$.}
\label{takeoff}
\end{table}
\endgroup

We observe again the need for a sizeable color-suppressed tree amplitude with a phase of about $-70^\circ$ relative to the color-favored tree amplitude.  In the last line of Table~\ref{PPpara}, the ratio $|C/T|$ has values $\agt 0.7$ that are not only at odds with the ratio of the effective Wilson coefficient $a_2$ to $a_1$, with a typical value of about 0.20~\cite{Beneke} in the QCDF calculations, but even larger than those found in previous analyses~\cite{cwPP}.  Comparing Schemes A and B or Schemes C and D, the ratio has a reduced central value when the $E$ and $P\!A$ amplitudes are included in the fits.  Such a large $|C|$ could be thought to be attributed to some particular set of observables, such as the $B^0 \to \pi^0\pi^0$ and/or $K^0 \pi^0$ decays.  To examine this idea, we have tried fits without observables of the $\pi^0\pi^0$ or $K^0\pi^0$ decays, where the $C$ amplitude plays an essential role, and compared them with Scheme B (to avoid complications from modes with $\eta$ or $\eta'$).  After trying new fits without some of the observables of the $\pi^0\pi^0$ and $K^0\pi^0$ modes, we see no reduction in $|C|$ and no significant difference in the other parameters, except for the phase $\delta_{P_{EW}}$, as shown in Table~\ref{takeoff}.  This implies that the large $|C|$ is required not just by any individual modes mentioned above. We shall see below that a large complex $C$ amplitude is a consequence of fitting to the observed direct CP asymmetries in $B\to K\pi$ decays.  We note in passing that the $P_{EW}$ amplitude has a strong phase of $\sim -80^\circ$ according to the global fits in Schemes C and D.  Such a large phase is unexpected within the perturbative formalism.  A similar phase is also found in the $P_{EW,V}$ amplitude for the $VP$ decays.

In the absence of the $c'$ amplitude, we see from Table~\ref{PPamplitude1} that the $K^+\pi^0$ and $K^+\pi^-$ decays are expected to have the same CP asymmetry.  However, experimentally $A_{CP}(K^+\pi^0)=0.040\pm0.021$ has a sign opposite to that of $A_{CP}(K^+ \pi^-)=-0.082\pm0.006$ (see Table~\ref{PPamplitude1}). This leads to the so-called $K\pi$ CP-puzzle; that is, $\Delta A_{K\pi}\equiv A_{CP}(K^+\pi^0)-A_{CP}(K^+\pi^-)=0.122\pm0.022$ shows a non-vanishing CP asymmetry difference at 5$\sigma$ level.  When the large complex amplitude $C$ is turned on, one can explicitly check that the sign of $A_{CP}(K^+\pi^0)$ is flipped and hence this basically resolves the $K\pi$ puzzle. Moreover, it helps solve the rate deficit problem with the $B^0\to\pi^0\pi^0$ decay.

One piece of evidence that one can take $\xi_p = 1$ comes from a comparison between $|p|$ and $|p'|$.  When the color-suppressed EW penguin amplitude is neglected, the ratio of them is equal to $|V_{cd}|/|V_{cs}|$ divided by $\xi_p$.  To obtain this ratio, we take the averaged amplitude of $B^+\to K^+\o K^0$ and $B^0\to K^0\o K^0$, both of which involve only the $p$ amplitude, and compare it with the amplitude obtained from $B^+\to K^0\pi^+$, which involves purely $p'$.  In the end, we find the ratio to be $0.23\pm0.01$, consistent with $|V_{cd}|/|V_{cs}| \simeq 0.23$.  Therefore, the flavor SU(3) breaking is negligible for penguin amplitudes.

The flavor-singlet amplitude plays an essential role particularly in explaining the branching fractions of the $\eta' K$ decays.  It is found to be $\sim 60\%$ of the QCD penguin amplitude and $\sim 4$ times larger than the EW penguin amplitude.  The associated phase is $\sim -100^\circ$ with respect to the $T$ amplitude.

It is noted that the fit quality of Scheme C is one order of magnitude worse than that of Scheme A.  However, the extracted parameters show sufficient consistency, with $|T|$, $|P|$, $|E|$, $|S|$ and their associated strong phases having high stability across the fits.  The strong phase of the EW penguin amplitude, $\delta_{P_{EW}}$, is most unstable when we go from restricted fits to global fits, with corresponding small changes in the magnitude and phase of $C$.  However, both the magnitude and strong phase of $P_{EW}$ are pretty stable within the restricted or global fits, independent of whether $E$ is included or not.

We have tried a fit with the $\eta$-$\eta'$ mixing angle $\phi$ as a free parameter.  It turns out that the data also favor a value around $46^\circ$ quoted in Ref.~\cite{phi}.  By modifying Schemes C and D to include $\phi$ as an additional parameter, for example, we obtain $\phi = (49^{+2}_{-5})^\circ$ and $(48^{+2}_{-4})^\circ$, respectively.  If we fix $\phi$ at the ``magic mixing angle'' of $35.3^\circ$, some observables will deviate a lot from measurements, notably the branching fractions of $B^+\to\eta K^+$ and $B^+\to\pi^+\pi^0$, and therefore result in an even higher $\chi^2_{\rm min}$.  We have also tried fits with $\xi_{t,c} = f_K/f_\pi$ and $\xi_{p,s} = 1$, but see no significant change in the fit quality.  We thus conclude that the flavor SU(3) symmetry in this sector is a sufficiently good working principle.

\subsection{Predictions}

\begingroup
\begin{table}[t!]
\begin{tabular}{lc|lc}
\hline\hline
\multicolumn{2}{c|}{$B^{+,0}$ decays}
& \multicolumn{2}{c}{$B_s$ decays} \\
~~Observable                  & ~~~~~Scheme B~~~~~    & ~~Observable   & ~~~~~Scheme B~~~~~ \\
\hline
$BF(\pi^+\pi^0)$          &$5.46\pm1.14$     & ~~$BF(K^+\pi^-)$         &$5.88\pm0.99$\\
$BF(K^+\o K^0)$         &$1.04\pm0.02$     & ~~$BF(\pi^0\o K^0)$    &$1.52\pm0.41$ \\
$BF(K^+K^-)$              &$0.13\pm0.06$     & ~~$BF(K^+K^-)$          &$18.89\pm3.35$ \\
$BF(K^0\o K^0)$         &$0.93\pm0.12$     & ~~$BF(K^0\o K^0)$      &$18.50\pm2.68$ \\
$BF(\pi^+\pi^-)$           &$5.16\pm1.28$     & ~~$BF(\pi^+\pi^-)$        &$0.67\pm0.61$   \\
$BF(\pi^0\pi^0)$          &$1.43\pm0.55$     & ~~$BF(\pi^0\pi^0)$        &$0.33\pm0.31$   \\
$BF(\pi^+K^0)$           &$23.55\pm0.41$   &&\\
$BF(\pi^0K^+)$           &$12.58\pm0.60$   &&\\
$BF(K^+\pi^-)$            &$20.20\pm0.39$   &&\\
$BF(K^0\pi^0)$           &$9.55\pm0.51$     &&\\
\hline
$A_{CP}(\pi^+\pi^0)$ &$-0.004\pm0.038$ & ~~$A_{CP}(K^+\pi^-)$  &$0.269\pm0.041$\\
$A_{CP}(K^+\o K^0)$&0                            & ~~${\cal A}(\pi^0K_S)$ &$0.635\pm0.124$ \\
$A_{CP}(K^+K^-)$     &$-0.182\pm0.787$ & ~~$A_{CP}(K^+K^-)$    &$-0.087\pm0.024$ \\
${\cal A}(K^0\o K^0)$ &$0.005\pm0.043$ & ~~${\cal A}(K^0\o K^0)$&$-0.072\pm0.039$\\
${\cal A}(\pi^+\pi^-)$   &$0.335\pm0.108$ & ~~${\cal A}(\pi^+\pi^-)$  &$0.036\pm0.155$ \\
${\cal A}(\pi^0\pi^0)$  &$0.354\pm0.192$ & ~~${\cal A}(\pi^0\pi^0)$  &$0.036\pm0.155$\\
$A_{CP}(K^0\pi^+)$   &0                           &&\\
$A_{CP}(K^+\pi^0)$   &$0.025\pm0.033$ &&\\
$A_{CP}(K^+\pi^-)$    &$-0.081\pm0.014$&&\\
${\cal A}(K_S\pi^0)$   &$-0.105\pm0.026$&&\\
          \hline
${\cal S}(K^0\o K^0)$  &$0.000\pm0.000$ & ~~${\cal S}(\pi^0K_S)$   &$-0.048\pm0.159$\\
${\cal S}(\pi^+\pi^-)$    &$-0.730\pm0.071$& ~~${\cal S}(K^+K^-)$     & $0.134\pm0.036$\\
${\cal S}(\pi^0\pi^0)$   &$0.791\pm0.138$ & ~~${\cal S}(K^0\o K^0)$ &$-0.039\pm0.001$\\
${\cal S}(K_S\pi^0)$    &$0.783\pm0.016$ & ~~${\cal S}(\pi^+\pi^-)$   &$0.120\pm0.088$ \\
                                    &                             & ~~${\cal S}(\pi^0\pi^0)$   &$0.120\pm0.190$\\
\hline\hline
\end{tabular}
\caption{Predictions based upon the theory parameters extracted in fit Scheme B in Table~\ref{PPpara}.  The left (right) two columns are for the $B^{+,0}$ ($B_s$) decays without involving the flavor-singlet amplitude.  All branching fractions are quoted in units of $10^{-6}$.}
\label{prenos}
\end{table}
\endgroup

Using the fit results obtained in the previous section, we predict the branching fractions and CP asymmetries of all the $PP$ decay modes.  Such predictions serve three purposes: (i) to see whether the fit results are compatible with individual measured observables, (ii) to compare with predictions made by perturbative approaches, and (iii) to test the working assumption of flavor SU(3) symmetry using future measurements of the yet observed ones, particularly those of the $B_s$ meson decays.  Our predictions based on Scheme B are given in Table~\ref{prenos} for all the $B$ decays without involving the flavor-singlet contribution.

It is noted that the CP asymmetries of some modes are predicted to be zero because they involve only a single flavor diagram in our analysis.  The uncertainty in ${\cal S}(B_s\to K^0\o K^0)$ comes purely from the errors in the CKM matrix elements, which we take to be zero, and is thus vanishing.

\begingroup
\squeezetable
\begin{table}[t!]
\begin{tabular}{lccccc}
\hline\hline
  ~~Observable~~ & ~~Data~~                       & ~~This Work~~ & ~~QCDF~~ & ~~pQCD~~ & ~~SCET~~ \\
         \hline
$BF(\pi^+\pi^0)$    &$5.48^{+0.35}_{-0.34}$   &$5.40\pm0.79$   &$5.9^{+2.2+1.4}_{-1.1-1.1}$ &$\sim 6.6$ \cite{Li:2014haa}                                    &$5.2\pm1.6\pm2.1\pm0.6$\\
$BF(K^+\o K^0)$   &$1.19\pm0.18$                &$1.03\pm0.02$  &$1.8^{+0.9+0.7}_{-0.5-0.5}$&1.66 \cite{KKPPp}                                                               &$1.1\pm0.4\pm1.4\pm0.03$\\
$BF(\e\pi^+)$         &$4.02\pm0.27$                &$3.88\pm0.39$  &$5.0^{+1.2+0.9}_{-0.6-0.7}$&$4.1^{+1.5}_{-1.1}$ \cite{PPpietap}                                    &$4.9\pm1.7\pm1.0\pm0.5$\\
$BF(\e'\pi^+)$        &$2.7^{+0.5}_{-0.4}$         &$5.59\pm0.54$   &$3.8^{+1.3+0.9}_{-0.6-0.6}$&$2.4^{+0.8}_{-0.5}\pm0.2\pm0.3$ \cite{PPpietap}              &$2.4\pm1.2\pm0.2\pm0.4$\\
$BF(K^+K^-)$         &$0.12\pm0.05$                &$0.15\pm0.05$   &$0.10^{+0.03}_{-0.02}\pm0.03$                  &0.046 \cite{KKPPp}                                                             & \\
$BF(K^0\o K^0)$   &$1.21\pm0.16$                &$0.89\pm0.11$   &$2.1^{+1.0+0.8}_{-0.6-0.6}$ &1.75 \cite{KKPPp}                                                              &$1.0\pm0.4\pm1.4\pm0.03$\\
$BF(\pi^+\pi^-)$     &$5.10\pm0.19$                &$5.17\pm1.03$   &$7.0^{+0.4}_{-0.7}\pm0.7$ &$\sim 6.4$ \cite{Li:2014haa}                                   &$5.4\pm1.3\pm1.4\pm0.4$\\
$BF(\pi^0\pi^0)$    &$1.17\pm0.13$                &$1.88\pm0.42$    &$1.1^{+1.0+0.7}_{-0.4-0.3}$ &$\sim 1.2$ \cite{Li:2014haa}                            &$0.84\pm0.29\pm0.30\pm0.19$\\
$BF(\e\pi^0)$         &$<1.5$                            &$0.56\pm0.03$    &$0.36^{+0.03+0.13}_{-0.02-0.10}$ &$0.23\pm0.08$ \cite{PPpietap}                                         &$0.88\pm0.54\pm0.06\pm0.42$\\
$BF(\e'\pi^0)$        &$1.2\pm0.4$                   &$1.21\pm0.16$     &$0.42^{+0.21+0.18}_{-0.09-0.12}$ &$0.19\pm0.02\pm0.03^{+0.04}_{-0.05}$ \cite{PPpietap}  &$2.3\pm0.8\pm0.3\pm2.7$\\
$BF(\e\e)$             &$<1.0$                            &$0.77\pm0.12$    &$0.32^{+0.13+0.07}_{-0.05-0.06}$ &$0.067^{+0.032}_{-0.025}$ \cite{PPeta2p}                        &$0.69\pm0.38\pm0.13\pm0.58$\\
$BF(\e'\e)$            &$<1.2$                            &$1.99\pm0.26$    &$0.36^{+0.24+0.12}_{-0.10-0.08}$ &$0.018\pm0.011$ \cite{PPeta2p}                                       &$1.0\pm0.5\pm0.1\pm1.5$\\
$BF(\e'\e')$           &$<1.7$                            &$1.60\pm0.20$    &$0.22^{+0.14+0.08}_{-0.06-0.06}$ &$0.011^{+0.012}_{-0.009}$ \cite{PPeta2p}                        &$0.57\pm0.23\pm0.03\pm0.69$\\
$BF(K^0\pi^+)$     &$23.79\pm0.75$             &$23.53\pm0.42$  &$21.7^{+9.2+9.0}_{-6.0-6.9}$ &$\sim 21.1$ \cite{Li:2014haa}                               &$20.8\pm7.9\pm0.6\pm0.7$\\
$BF(K^+\pi^0)$     &$12.94^{+0.52}_{-0.51}$&$12.71\pm1.05$  &$12.5^{+4.7+4.9}_{-3.0-3.8}$ &$\sim 12.9$ \cite{Li:2014haa}                                &$11.3\pm4.1\pm1.0\pm0.3$\\
$BF(\e K^+)$        &$2.36^{+0.22}_{-0.21}$   &$1.93\pm0.31$   &$2.2^{+1.7+1.1}_{-1.0-0.9}$      \cite{etaQ}                &$3.2^{+3.2}_{-1.8}$ \cite{PPetap}                                         &$2.7\pm4.8\pm0.4\pm0.3$\\
$BF(\e'K^+)$        &$71.1\pm2.6$                  &$70.92\pm8.54$ &$74.5^{+57.9+25.6}_{-25.3-19.0}$   \cite{etaQ}        &$51.0^{+18.0}_{-10.9}$ \cite{PPetap}                                  &$69.5\pm27.0\pm4.3\pm7.7$\\
$BF(K^+\pi^-)$     &$19.57^{+0.53}_{-0.52}$ &$20.18\pm0.39$ &$19.3^{+7.9+8.2}_{-4.8-6.2}$ &$\sim 17.7$\cite{Li:2014haa}                                &$20.1\pm7.4\pm1.3\pm0.6$\\
$BF(K^0\pi^0)$     &$9.93\pm0.49$                &$9.73\pm0.82$   &$8.6^{+3.8+3.8}_{-2.2-2.9}$ &$\sim 7.2$ \cite{Li:2014haa}                                    &$9.4\pm3.6\pm0.2\pm0.3$\\
$BF(\e K^0)$         &$1.23^{+0.27}_{-0.24}$   &$1.49\pm0.27$   &$1.5^{+1.4+0.9}_{-0.8-0.7}$  \cite{etaQ}                   &$2.1^{+2.6}_{-1.5}$ \cite{PPetap}                                        &$2.4\pm4.4\pm0.2\pm0.3$\\
$BF(\e' K^0)$        &$66.1\pm3.1$                 &$66.51\pm7.97$  &$70.9^{+54.1+24.2}_{-23.8-18.0}$ \cite{etaQ}         &$50.3^{+16.8}_{-10.6}$ \cite{PPetap}                                  &$63.2\pm24.7\pm4.2\pm8.1$\\
\hline\hline
\end{tabular}\\
\caption{Predicted branching fractions in units of $10^{-6}$ for the $B^{0,+}$ decays based on Scheme D.  Unless otherwise noted, QCDF predictions are taken from Refs.~\cite{Q,etaQ}  and SCET predictions from Ref.~\cite{PPSC}. The pQCD predictions taken from Ref.~\cite{Li:2014haa} are for $S_e=-\pi/2$ with $S_e$ being a strong phase induced by Glauber gluons.}
\label{preglobal0Br}
\end{table}
\endgroup

\begingroup
\squeezetable
\begin{table}[t!]
\begin{tabular}{lccccc}
\hline\hline
  ~~Observable~~     & ~~Data~~             & ~~This Work~~   & ~~QCDF~~ & ~~pQCD~~ & ~~SCET~~ \\
\hline
$A_{CP}(\pi^+\pi^0)$&$0.026\pm0.039$  &$0.069\pm0.027$ &$-0.0011\pm0.0001^{+0.0006}_{-0.0003}$ &$\sim -0.012$ \cite{Li:2014haa}                                                         &$<0.04$\\
$A_{CP}(K^+K_S)$  &$-0.086\pm0.100$ &0                           &$-0.064^{+0.008}_{-0.006}\pm0.018$                   &0.11 \cite{KKPPp}                                                                              &-\\
$A_{CP}(\e\pi^+)$     &$-0.14\pm0.05$    &$-0.081\pm0.074$&$-0.050^{+0.024+0.084}_{-0.034-0.103}$ &$-0.37^{+0.09}_{-0.07}$ \cite{PPpietap}                                            &$0.05\pm0.19\pm0.21\pm0.05$\\
$A_{CP}(\e'\pi^+)$    &$0.06\pm0.15$      &$0.374\pm0.087$  &$0.016^{+0.050+0.094}_{-0.082-0.111}$ &$-0.33^{+0.07}_{-0.08}$ \cite{PPpietap}                                             &$0.21\pm0.12\pm0.10\pm0.14$\\
$A_{CP}(K^+K^-)$    & -                           & $0.004\pm0.612$  & 0                                                                           &0.29 \cite{KKPPp}                                                                               &-\\
${\cal A}(K^0\o K^0)$&$0.06\pm0.26$     & $0.017\pm0.041$  &$-0.100\pm0.007^{+0.010}_{-0.019}$ &0 \cite{KKPPp}                                                                                    &-\\
${\cal A}(\pi^+\pi^-)$  &$0.31\pm0.05$     &$0.326\pm0.081$   &$0.170^{+0.013+0.043}_{-0.012-0.087}$ &$\sim 0.17$ \cite{Li:2014haa}                                             &$0.20\pm0.17\pm0.19\pm0.05$\\
${\cal A}(\pi^0\pi^0)$   &$0.03\pm0.17$      &$0.611\pm0.113$   &$0.572^{+0.148+0.303}_{-0.208-0.346}$ &$\sim 0.36$ \cite{Li:2014haa}                 &$-0.58\pm0.39\pm0.39\pm0.13$\\
${\cal A}(\e\pi^0)$     &-                             &$0.566\pm0.114$   &$-0.052^{+0.028+0.246}_{-0.050-0.156}$ &$-0.42^{+0.10}_{-0.13}$ \cite{PPpietap}                                              &$0.03\pm0.10\pm0.12\pm0.05$\\
${\cal A}(\e'\pi^0)$     &-                            &$0.385\pm0.114$   &$-0.073^{+0.010+0.176}_{-0.018-0.140}$ &$-0.36^{+0.11}_{-0.10}$ \cite{PPpietap}                                               &$-0.24\pm0.10\pm0.19\pm0.24$\\
${\cal A}(\e\e)$          &-                            &$-0.405\pm0.129$  &$-0.635^{+0.104+0.098}_{-0.064-0.124}$ &$-0.33^{+0.026+0.041+0.035}_{-0.028-0.038-0.000}$ \cite{PPeta2p}&$-0.09\pm0.24\pm0.21\pm0.04$\\
${\cal A}(\e\e')$         &-                            &$-0.394\pm0.117$    &$-0.592^{+0.072+0.038}_{-0.068-0.048}$ &$0.774^{+0.000+0.069+0.080}_{-0.056-0.112-0.090}$ \cite{PPeta2p}&-\\
${\cal A}(\e'\e')$         &-                            &$-0.122\pm0.136$  &$-0.449\pm0.031^{+0.085}_{-0.092}$ &$0.237^{+0.100+0.185+0.060}_{-0.069-0.169-0.085}$ \cite{PPeta2p}&-\\
$A_{CP}(K_S\pi^+)$&$-0.017\pm0.016$ &0                             &$0.0028\pm0.0003^{+0.0009}_{-0.0010}$              &$\sim 0.001$ \cite{Li:2014haa}                                                                                                            &$<0.05$\\
$A_{CP}(K^+\pi^0)$ &$0.040\pm0.021$  &$0.047\pm0.025$   &$0.049^{+0.039+0.044}_{-0.021-0.054}$ &$\sim 0.10$ \cite{Li:2014haa}                                               &$-0.11\pm0.09\pm0.11\pm0.02$\\
$A_{CP}(\e K^+)$    &$-0.37\pm0.08$     &$-0.426\pm0.043$  &$-0.145^{+0.103+0.155}_{-0.260-0.107}$ \cite{etaQ}        &$-0.117^{+0.068+0.039+0.029}_{-0.096-0.042-0.056}$ \cite{PPetap}   &$0.33\pm0.30\pm0.07\pm0.03$\\
$A_{CP}(\e' K^+)$   &$0.013\pm0.017$   &$-0.027\pm0.008$  &$0.0045^{+0.0069+0.0120}_{-0.0055-0.0098}$ \cite{etaQ} &$-0.062^{+0.012+0.013+0.013}_{-0.011-0.010-0.010}$ \cite{PPetap}   &$-0.010\pm0.006\pm0.007\pm0.005$\\
$A_{CP}(K^+\pi^-)$ &$-0.082\pm0.006$  &$-0.080\pm0.011$  &$-0.074^{+0.017+0.043}_{-0.015-0.048}$ &$\sim -0.11$\cite{Li:2014haa}                                               &$-0.06\pm0.05\pm0.06\pm0.02$\\
${\cal A}(K_S\pi^0)$   &$-0.01\pm0.10$   &$-0.173\pm0.019$  &$-0.106^{+0.027+0.056}_{-0.038-0.043}$ &$\sim -0.21$\cite{Li:2014haa}                                                             &$0.05\pm0.04\pm0.04\pm0.01$\\
${\cal A}(\e K_S)$       &-                          &$-0.301\pm0.041$  &$-0.236^{+0.098+0.126}_{-0.262-0.125}$  \cite{etaQ}          &$-0.127\pm0.041^{+0.032+0.032}_{-0.015-0.067}$ \cite{PPetap}          &$0.21\pm0.20\pm0.04\pm0.03$\\
${\cal A}(\e' K_S)$      &$0.05\pm0.04$   &$0.022\pm0.006$   &$0.030^{+0.006}_{-0.005}\pm0.008$  \cite{etaQ} &$0.023^{+0.005+0.003+0.002}_{-0.004-0.006-0.001}$ \cite{PPetap}&$0.011\pm0.006\pm0.012\pm0.002$\\
           \hline
${\cal S}(K^0\o{K^0})$      &$-1.08\pm0.49$     &0                             &      -                                                                                     &                  -                                                                                               &-\\
${\cal S}(\pi^+\pi^-)$         &$-0.66\pm0.06$     &$-0.717\pm0.061$  &$-0.69^{+0.08+0.19}_{-0.10-0.09}$ &$\sim -0.43$ \cite{Li:2014haa}                                                  &$-0.86\pm0.07\pm0.07\pm0.02$\\
${\cal S}(\pi^0\pi^0)$        &-                             &$0.454\pm0.112$   &          -                                                                                  &$\sim 0.63$ \cite{Li:2014haa}                                                                       &$0.71\pm0.34\pm0.33\pm0.10$\\
${\cal S}(\e\pi^0)$            &-                              &$-0.098\pm0.338$ &$0.08^{+0.06+0.19}_{-0.12-0.23}$ &$0.67^{+0.05}_{-0.11}$ \cite{PPpietap}                                                     &$-0.90\pm0.08\pm0.03\pm0.22$\\
${\cal S}(\e'\pi^0)$           &-                              &$0.142\pm0.234$ &$0.16^{+0.05+0.11}_{-0.07-0.14}$ &$0.67^{+0.05}_{-0.11}$ \cite{PPpietap}                                                     &$-0.96\pm0.03\pm0.05\pm0.11$\\
${\cal S}(\e\e)$                &-                              &$-0.796\pm0.077$ &$-0.77^{+0.07+0.12}_{-0.05-0.06}$ &$0.535^{+0.000+0.031+0.021}_{-0.034-0.027-0.001}$ \cite{PPeta2p}    &$-0.98\pm0.06\pm0.03\pm0.09$\\
${\cal S}(\e'\e)$               &-                              &$-0.903\pm0.049$ &$-0.76^{+0.07+0.06}_{-0.05-0.03}$ &$-0.131^{+0.547+0.090+0.100}_{-0.488-0.099-0.062}$ \cite{PPeta2p}    &$-0.82\pm0.02\pm0.04\pm0.77$\\
${\cal S}(\e'\e')$              &-                              &$-0.964\pm0.037$  &$-0.85^{+0.03+0.07}_{-0.02-0.06}$ &$0.932^{+0.049+0.052+0.022}_{-0.024-0.111-0.021}$ \cite{PPeta2p}      &$-0.59\pm0.05\pm0.08\pm1.10$\\
${\cal S}(K_S\pi^0)$        &$0.57\pm0.17$      &$0.754\pm0.014$   &$0.79^{+0.06}_{-0.04}\pm0.04$ &$\sim 0.69$ \cite{Li:2014haa}                                                     &$0.80\pm0.02\pm0.02\pm0.01$\\
${\cal S}(\e K_S)$           &-                             &$0.592\pm0.035$   &$0.79^{+0.04+0.08}_{-0.06-0.06}$ &$0.619^{+0.358+0.353}_{-0.650-0.643}$ \cite{PPetap}                              &$0.69\pm0.15\pm0.05\pm0.01$\\
${\cal S}(\e' K_S)$          &$0.63\pm0.06$     &$0.685\pm0.004$   & $0.67\pm0.01\pm0.01$ &$0.627^{+0.355+0.354}_{-0.650-0.647}$ \cite{PPetap}                              &$0.706\pm0.005\pm0.006\pm0.003$\\
\hline\hline
\end{tabular}\\
\caption{Same as Table~\ref{preglobal0Br} but for CP asymmetries.
}
\label{preglobal0Acp}
\end{table}
\endgroup

\begingroup
\squeezetable
\begin{table}[t!p]
\begin{tabular}{lccccc}
\hline\hline
  ~~Observable~~    & ~~Data~~        & ~~This Work~~ & ~~QCDF~~ & ~~pQCD~~ & ~~SCET~~ \\
\hline
$BF(\pi^+K^-)$          &$5.4\pm0.6$     &$5.86\pm0.78$   &$5.3^{+0.4+0.4}_{-0.8-0.5}$                  &$7.6^{+3.2}_{-2.3}\pm0.7\pm0.5$                                &$4.9\pm1.2\pm1.3\pm0.3$\\
$BF(\pi^0\o K^0)$     &-                        &$2.25\pm0.33$   &$1.7^{+2.5+1.2}_{-0.8-0.5}$                  &$0.16^{+0.05+0.10+0.02}_{-0.04-0.05-0.01}$              &$0.76\pm0.26\pm0.27\pm0.17$\\
$BF(\e\o K^0)$          &-                        &$0.97\pm0.16$ &$0.75^{+1.10+0.51}_{-0.35-0.22}$         &$0.11^{+0.05+0.06}_{-0.03-0.03}\pm0.01$                  &$0.80\pm0.48\pm0.29\pm0.18$\\
$BF(\e'\o K^0)$         &-                        &$3.94\pm0.39$  &$2.8^{+2.5+1.1}_{-1.0-0.8}$                   &$0.72^{+0.20+0.28+0.11}_{-0.16-0.17-0.05}$              &$4.5\pm1.5\pm0.4\pm0.5$\\
$BF(K^+K^-)$           &$24.5\pm1.8$   &$17.90\pm2.98$ &$25.2^{+12.7+12.5}_{-7.2-9.1}$             &$13.6^{+4.2+7.5+0.7}_{-3.2-4.1-0.2}$                          &$18.2\pm6.7\pm1.1\pm0.5$\\
$BF(K^0\o K^0)$      &$<66$               &$17.48\pm2.36$ &$26.1^{+13.5+12.9}_{-8.1-9.4}$             &$15.6^{+5.0+8.3+0.0}_{-3.8-4.7-0.0}$                          &$17.7\pm6.6\pm0.5\pm0.6$\\
$BF(\pi^+\pi^-)$        &$0.73\pm0.14$ &$0.80\pm0.55$   &$0.26\pm0.00^{+0.10}_{-0.09}$              &$0.57^{+0.16+0.09+0.01}_{-0.13-0.10-0.00}$              &-\\
$BF(\pi^0\pi^0)$       &-                        &$0.40\pm0.27$   &$0.13\pm0.0\pm0.05$                             &$0.28^{+0.08+0.04+0.01}_{-0.07-0.05-0.00}$              &-\\
$BF(\e\pi^0)$           &-                        &$0.12\pm0.07$   &$0.05^{+0.03+0.02}_{-0.01-0.01}$          &$0.05\pm0.02\pm0.01\pm0.00$                                    &$0.014\pm0.004\pm0.005\pm0.004$\\
$BF(\e'\pi^0)$          &-                        &$0.12\pm0.06$    &$0.04^{+0.01+0.01}_{-0.00-0.00}$          &$0.11^{+0.05+0.02}_{-0.03-0.01}\pm0.00$                   &$0.006\pm0.003\pm0.002^{+0.064}_{-0.006}$\\
$BF(\e\e)$               &-                        &$8.24\pm1.53$  &$10.9^{+6.3+5.7}_{-4.0-4.2}$                  &$8.0^{+2.6+4.7}_{-1.9-2.5}\pm0.0$                               &$7.1\pm6.4\pm0.2\pm0.8$\\
$BF(\e\e')$              &-                        &$33.47\pm3.64$  &$41.2^{+27.3+17.8}_{-12.9-13.1}$          &$21.0^{+6.0+10.0}_{-4.6-5.6}\pm0.0$                            &$24.0\pm13.6\pm1.4\pm2.7$\\
$BF(\e'\e')$              &-                        &$41.48\pm6.25$  &$47.9^{+41.6+20.9}_{-17.1-15.3}$          &$14.0^{+3.2+6.2}_{-2.7-3.9}\pm0.0$                              &$44.3\pm19.7\pm2.3\pm17.1$\\
\hline
$A_{CP}(\pi^+K^-)$  &$0.26\pm0.04$ &$0.266\pm0.033$ &$0.207^{+0.050+0.039}_{-0.030-0.088}$&$0.241^{+0.039+0.033+0.023}_{-0.036-0.030-0.012}$  &$0.20\pm0.17\pm0.19\pm0.05$\\
${\cal A}(\pi^0K_S)$  &-                        &$0.724\pm0.054$ &$0.363^{+0.174+0.266}_{-0.182-0.243}$&$0.594^{+0.018+0.074+0.022}_{-0.040-0.113-0.035}$  &$-0.58\pm0.39\pm0.39\pm0.13$\\
${\cal A}(\e K_S)$     &-                        &$0.452\pm0.057$ &$0.334^{+0.228+0.257}_{-0.238-0.216}$&$0.564^{+0.029+0.068+0.031}_{-0.034-0.080-0.034}$  &$-0.56\pm0.46\pm0.14\pm0.06$\\
${\cal A}(\e' K_S)$     &-                        &$-0.367\pm0.089$&$-0.493^{+0.062+0.160}_{-0.050-0.130}$&$-0.199^{+0.016+0.051+0.014}
_{-0.014-0.050-0.009}$&$-0.14\pm0.07\pm0.16\pm0.02$\\
${\cal A}(K^+K^-)$     &$-0.14\pm0.11$&$-0.090\pm0.021$&$-0.077^{+0.016+0.040}_{-0.012-0.051}$&$-0.233
^{+0.009+0.049+0.008}_{-0.002-0.044-0.011}$&$-0.06\pm0.05\pm0.06\pm0.02$\\
${\cal A}(K^0\o K^0)$&-                       &$-0.075\pm0.035$ &$0.0040\pm0.0004^{+0.0010}_{-0.0004}$&$0$                                                                                &$<0.1$\\
${\cal A}(\pi^+\pi^-)$  &-                       &$-0.001\pm0.110$ &$0$                                                           &$-0.012^{+0.001}_{-0.004}\pm0.012\pm0.001$             &-\\
${\cal A}(\pi^0\pi^0)$ &-                       &$-0.001\pm0.110$ &$0$                                                            &$-0.012^{+0.001}_{-0.004}\pm0.012\pm0.001$            &-\\
${\cal A}(\e\pi^0)$     &-                       &$-0.165\pm0.292$  &$0.961^{+0.016+0.018}_{-0.143-0.371}$  &$-0.004^{+0.006}_{-0.007}\pm0.022\pm0.000$            &-\\
${\cal A}(\e'\pi^0)$    &-                       &$0.259\pm0.335$ &$0.429^{+0.023+0.310}_{-0.081-0.409}$  &$0.206^{+0.000+0.020+0.028}_{-0.007-0.025-0.012}$&-\\
${\cal A}(\e\e)$         &-                       &$-0.116\pm0.018$ &$-0.050^{+0.015+0.038}_{-0.025-0.028}$ &$-0.006\pm0.002^{+0.006+0.000}_{-0.005-0.001}$      &$0.079\pm0.049\pm0.027\pm0.015$\\
${\cal A}(\e\e')$        &-                       &$-0.009\pm0.003$ &$-0.006^{+0.003+0.005}_{-0.004-0.003}$  &$-0.013\pm0.000^{+0.001}_{-0.002}\pm0.001$            &$0.0004\pm0.0014\pm0.0039\pm0.0043$\\
${\cal A}(\e'\e')$       &-                       &$0.016\pm0.009$  &$0.032^{+0.008+0.010}_{-0.006-0.012}$   &$0.019\pm0.002^{+0.003+0.002}_{-0.004-0.001}$        &$0.009\pm0.004\pm0.006\pm0.019$\\
          \hline
${\cal S}(\pi^0K_S)$        &-                       &$0.302\pm0.080$  &$0.08^{+0.29+0.23}_{-0.27-0.26}$            &$-0.61^{+0.08+0.23+0.01}_{-0.06-0.19-0.03}$               &$-0.16\pm0.41\pm0.33\pm0.17$\\
${\cal S}(\e K_S)$           &-                       &$0.787\pm0.042$ &$0.26^{+0.33+0.21}_{-0.44-0.30}$             &$-0.43^{+0.03+0.22+0.02}_{-0.04-0.21-0.03}$               &$0.82\pm0.32\pm0.11\pm0.04$\\
${\cal S}(\e' K_S)$          &-                       &$0.191\pm0.090$  &$0.08^{+0.21+0.20}_{-0.17-0.16}$             &$-0.68^{+0.01+0.06}_{-0.02-0.05}\pm0.00$                    &$0.38\pm0.08\pm0.10\pm0.04$\\
${\cal S}(K^+K^-)$          &$0.30\pm0.13$& $0.140\pm0.030$ &$0.22^{+0.04+0.05}_{-0.05-0.03}$             &$0.28\pm0.03\pm0.04^{+0.02}_{-0.01}$                          &$0.19\pm0.04\pm0.04\pm0.01$\\
${\cal S}(K^0\o K^0)$     &-                       &$-0.039\pm0.001$ &$0.004\pm0.0^{+0.002}_{-0.001}$             &$0.04$                                                                              &-\\
${\cal S}(\pi^+\pi^-)$       &-                       &$0.114\pm0.061$ &$0.15\pm0.00\pm0$                                   &$0.14^{+0.02+0.08+0.09}_{-0.00-0.02-0.05}$                 &-\\
${\cal S}(\pi^0\pi^0)$      &-                       &$0.114\pm0.061$ &$0.15\pm0.00\pm0$                                   &$0.14^{+0.02+0.08+0.09}_{-0.00-0.02-0.05}$                 &-\\
${\cal S}(\e\pi^0)$          &-                        &$0.836\pm0.198$ &$0.26^{+0.06+0.48}_{-0.23-0.47}$             &$0.17\pm0.04^{+0.10}_{-0.12}\pm0.01$                          &$0.45\pm0.14\pm0.42\pm0.30$\\
${\cal S}(\e'\pi^0)$         &-                        &$0.953\pm0.116$ &$0.88^{+0.03+0.04}_{-0.15-0.29}$             &$-0.17^{+0.00+0.07+0.03}_{-0.01-0.08-0.05}$                 &-\\
${\cal S}(\e\e)$              &-                        &$-0.095\pm0.020$ &$-0.07^{+0.03+0.04}_{-0.06-0.05}$            &$0.03\pm0.00\pm0.01\pm0.00$                                        &$-0.026\pm0.040\pm0.030\pm0.014$\\
${\cal S}(\e\e')$             &-                        &$-0.036\pm0.007$&$-0.01^{+0.00}_{-0.01}\pm0.00$               &$0.04\pm0.00\pm0.00\pm0.00$                                        &$0.041\pm0.004\pm0.002\pm0.051$\\
${\cal S}(\e'\e')$            &-                        &$0.028\pm0.009$  &$0.04\pm0.01\pm0.01$                               &$0.04\pm0.00\pm0.01\pm0.00$                                        &$0.049\pm0.005\pm0.005\pm0.031$\\
\hline\hline
\end{tabular}\\
\caption{Predicted results for the $B_s$ decays based on Scheme D. QCDF predictions are taken from Ref.~\cite{sQ},  pQCD predictions from Ref.~\cite{sp}, and SCET predictions from Ref.~\cite{PPSC}.  Branching fractions are quoted in units of $10^{-6}$. }
\label{preglobal1}
\end{table}
\endgroup

For global fits in the $PP$ sector, we choose to present the predictions based on Scheme D in Tables~\ref{preglobal0Br}--\ref{preglobal1}. Table \ref{preglobal0Br} lists the branching fractions of all the $B^{0,+}$ decays, Table~\ref{preglobal0Acp} the CP asymmetries of all the $B^{0,+}$ decays, and Table~\ref{preglobal1} all the observables for the $B_s$ decays.  In all the tables, we also list available experimental data and predictions made by QCDF, pQCD, and SCET.  In the following, we discuss those observables with large discrepancies between our prediction and data or other approaches.

As seen in Table~\ref{preglobal0Br}, our prediction for $BF(\eta'\pi^+)$ is roughly twice larger than the measured value and most other perturbative calculations.  This is because with the choice of $\phi = 46^\circ$, there is constructive interference between the flavor-singlet diagram and the others in the $\eta\pi^+$ and $\eta'\pi^+$ decays.  Moreover, the flavor-singlet component of the latter is bigger than the former.  Therefore, it is expected that the latter has an even larger branching fraction than the former.  It is noted that there is a significant difference, characterized by the scale factor of $1.36$, for this observable among BaBar, Belle, and CLEO.

It is a well-known problem that the branching fraction of $B^0 \to \pi^0\pi^0$ used to be significantly larger than most perturbative calculations.\footnote{It is known that there is a huge cancelation between the vertex and na{\"i}vely factorizable terms so that the real part of the $C$ amplitude is governed by spectator interactions, while its imaginary part comes mainly from the vertex corrections~\cite{QCDF}.  Based on this observation, recently there were two attempts trying to solve the $B^0 \to \pi^0\pi^0$ puzzle by enhancing the spectator contribution to $C$: one of them is to consider the Glauber gluon effects in the spectator amplitudes~\cite{Li:2014haa}, and the other argued that the renormalization scale for hard spectator interactions is significantly lower after applying the principle of maximum conformality~\cite{Qiao:2014lwa}.}
A preliminary Belle measurement of $B\!F(B^0\to\pi^0\pi^0)=(0.90\pm0.12\pm0.10)\times 10^{-6}$ \cite{pi0pi0exp}
brings it closer to the estimates made by QCDF and SCET, although the weighted average has the largest scale factor in the $PP$ sector.  Our predictions are about 20\% and 60\% larger in Schemes B and D, respectively.  As alluded to before, this is due to a large $|C|$ demanded by other observables.

The measured branching fraction of $B^0 \to \eta'\pi^0$ is much larger than the predictions made by QCDF and pQCD.  It can be nicely explained within our approach due to constructive interference between the QCD penguin and flavor-singlet diagrams, which subtend a phase less than $90^\circ$.

In Table~\ref{preglobal0Acp}, the measured value of $A_{CP}(\eta'\pi^+)$ and all predictions show a diversity, with pQCD having an opposite sign from  the others.  Our prediction of ${\cal A}(\pi^+\pi^-)$ agrees better with data, whereas the others tend to be smaller by at least 30\%.  The recently updated ${\cal A}(\pi^0\pi^0)$ has a scale factor of $1.94$ and is significantly different from all theory predictions.  We have a prediction for $A_{CP}(\eta K^+)$ very close to the measured value, established at $\sim 4.6\sigma$ level, while all the perturbative approaches have far-off central values.  Finally, theory predictions for ${\cal A}(\eta\eta')$, $A_{CP}(\eta'K^+)$, ${\cal A}(\eta K_S)$, ${\cal S}(\eta\pi^0)$, ${\cal S}(\eta'\pi^0)$, ${\cal S}(\eta\eta)$, and ${\cal S}(\eta'\eta')$ are quite different, and awaits more precise measurements to determine which one is favored.

With reference to Table~\ref{preglobal1} for $B_s$ decays, our predictions for $BF(\pi^+K^-)$ and $BF(K^+K^-)$ agree well with the measured values, though the predicted central value for the latter is slightly smaller.  The measured CP asymmetries are consistent with our predictions within errors.  Note that the $B_s \to \pi^+\pi^-$ and $\pi^0\pi^0$ decays are dominated by the penguin annihilation contribution.  Although the $P\!A$ amplitude is suppressed by one order of magnitude with respect to the $E$ amplitude (see Table~\ref{PPpara}), the CKM factors ($|Y_{sb}^c| \gg |Y_{sb}^u|$) render $|pa'| > |e'|$.  Our prediction $BF(B_s \to \pi^+\pi^-)=(0.80\pm0.55)\times 10^{-6}$ is in good agreement with the measured value of $(0.73 \pm 0.14) \times 10^{-6}$.  A related prediction is $BF(B_s \to \pi^0\pi^0) = (0.40 \pm 0.27) \times 10^{-6}$.  Note that it has been claimed in the literature that large flavor symmetry breaking effects between the annihilation amplitudes of $B_s$ and $B_{u,d}$ decays are needed in order to explain the data of $B_s\to \pi^+\pi^-$ and $B_d\to K^+K^-$~\cite{Wang:2013fya}.  This is not the case in the present work.

The pQCD approach gives much smaller branching fraction predictions in the $B_s \to \pi^0\o K^0$, $\eta \o K^0$, $\eta'\o K^0$ modes in comparison with the others.  The $\eta\eta'$ and $\eta'\eta'$ modes are predicted by us to have the largest branching fractions among the $B_s$ decays, whereas pQCD gives somewhat lower values for both.  As to the CP asymmetries, the following ones show significant disagreements among theory predictions: ${\cal A}(\eta\pi^0)$, ${\cal S}(\pi^0 K_S)$, ${\cal S}(\eta K_S)$, ${\cal S}(\eta' K_S)$, ${\cal S}(K^0 \o K^0)$, ${\cal S}(\eta \pi^0)$, and ${\cal S}(\eta' \pi^0)$.  In particular, our predictions for ${\cal S}(\eta K_S)$, ${\cal S}(\eta \pi^0)$, and ${\cal S}(\eta' \pi^0)$ are close to 1, whereas most others are smaller.  As far as the central values are concerned, our predictions for $BF(\eta\pi^0)$ and $BF(\eta'\pi^0)$ are roughly the same because of $\phi = 46^\circ$ and are larger than most other perturbative calculations because they are dominated by the $C$ amplitude.

\section{The $B\to VP$ Sector \label{sec:VP}}

\begingroup
\squeezetable
\begin{table}[t!h]
\begin{tabular}{ccccc}
\hline\hline
\multicolumn{2}{c}{Mode} & Flavor Amplitude & ~~~$BF$~~~ & ~~~$A_{CP}$~~~   \\
\hline
$B^+\to$ & $\o K^{*0}K^+$ & $p_P$                                                                                             & $<1.1$                      &-\\
&$K^{*+}\o K^0$ & $p_V$                                                                                                             & -                                &-\\
&$\r^0\p^+$        & $-\frac{1}{\sqrt2}(t_V+c_P+p_V-p_P)$                                                            & $8.3^{+1.2}_{-1.3}$  &$0.18^{+0.09*}_{-0.17}$\\
&$\r^+\p^0$        & $-\frac{1}{\sqrt2}(t_P+c_V+p_P-p_V)$                                                            & $10.9^{+1.4}_{-1.5}$&$0.02\pm0.11$\\
&$\r^+\e$            & $\frac{c_\phi}{\sqrt2}[t_P+c_V+p_P+p_V+(-\sqrt2t_\phi+2)s_V]$                  & $6.9\pm1.0$ (2.06)  &$0.11\pm0.11$\\
&$\r^+\e'$           & $\frac{s_\phi}{\sqrt2}[t_P+c_V+p_P+p_V+(\frac{\sqrt2}{t_\phi}+2)s_V]$         & $9.8^{+2.1}_{-2.0}$  &$0.26\pm0.17$\\
&$\w\p^+$          & $\frac{1}{\sqrt2}(t_V+c_P+p_P+p_V+2s_P)$                                                   & $6.9\pm0.5$            &$-0.02\pm0.06$\\
&$\f\p^+$           & $s_P$                                                                                                               & $<0.15$   \cite{phiKphipi}                   &-\\
\hline
$B^0\to$ &$\o K^{*0}K^0$ & $p_P$                                                                                              & -                                &-\\
&$K^{*0}\o K^0$ & $p_V$                                                                                                             & $<1.9$                      &-\\
&$\r^-\p^+$         & $-(t_V+p_V+e_P)$                                                                                          &$8.4\pm1.1$             &$-0.07\pm0.09$\\
&                         &                                                                                                                         &                                  &$0.05\pm0.08$\\
&$\r^+\p^-$         & $-(t_P+p_P+e_V)$                                                                                          &$14.6\pm1.6$          &$0.13\pm0.06$\\
&                         &                                                                                                                         &                                &$0.07\pm0.14$\\
&$\r^0\p^0$        & $-\frac{1}{2}(c_P+c_V-p_P-p_V-e_P-e_V)$                                                    & $2.0\pm0.5$ (1.05)  &$-0.27\pm0.24$\\
&                         &                                                                                                                         &                                  &$-0.23\pm0.34$\\
&$\r^0\e$            & $-\frac{c_\phi}{2}[c_P-c_V-p_P-p_V+(\sqrt2t_\phi-2)s_V+e_P+e_V]$              & $<1.5$                     &-\\
&$\r^0\e'$           & $-\frac{s_\phi}{2}[c_P-c_V-p_P-p_V+(-\frac{\sqrt2}{t_\phi}-2)s_V+e_P+e_V]$& $<1.3$                     &-\\
&$\w\p^0$          & $\frac{1}{2}(c_P-c_V+p_P+p_V+2s_P-e_P-e_V)$                                           & $<0.5$                     &-\\
&$\w\e$             &$\frac{c_\phi}{2}[c_P+c_V+p_P+p_V+2s_P+(-\sqrt2t_\phi+2)s_V+e_P+e_V]$& $<1.4$                     &-\\
&$\w\e'$            & $\frac{s_\phi}{2}[c_P+c_V+p_P+p_V+2s_P+(\frac{\sqrt2}{t_\phi}+2)s_V+e_P+e_V]$& $<1.8$                     &-\\
&$\f\p^0$          & $\frac{1}{\sqrt2}s_P$                                                                                         & $<0.15$                   &-\\
&$\f\e$              & $\frac{c_\phi}{\sqrt2}s_P$                                                                             & $<0.5$                     &-\\
&$\f\e'$             & $\frac{s_\phi}{\sqrt2}s_P$                                                                               & $<0.5$                     &-\\
&$K^{*-}K^+$   & $-e_P$                                                                               & -                   &-\\
&$K^{*+}K^-$   & $-e_V$                                                                                 &-                   &-\\
&\qquad$K^{*\pm}K^\mp$   & 
&$<0.4$ \cite{BsVPexp}                    &-\\
\hline
$B^0_s\to$ &$\o K^{*0}\p^0$& $-\frac{1}{\sqrt2}(c_V-p_V)$& - &-\\
&$K^{*-}\p^+$   & $-(t_V+p_V)$                                                                                                    & $3.3\pm1.2^*$ \cite{BsVPexp} &-\\
&$\r^+K^-$        & $-(t_P+p_P)$ & - &-\\
&$\r^0\o K^0$   & $-\frac{1}{\sqrt2}(c_P-p_P)$ & - &-\\
&$\o K^{*0}\e$  & $\frac{c_\phi}{\sqrt2}[c_V-\sqrt2t_\phi p_P+p_V+(-\sqrt2t_\phi+2)s_V]$& - &-\\
&$\o K^{*0}\e'$ & $\frac{s_\phi}{\sqrt2}[c_V+\frac{\sqrt2}{t_\phi}p_P+p_V+(\frac{\sqrt2}{t_\phi}+2)s_V]$& - &-\\
&$\w\o K^0$     & $\frac{1}{\sqrt2}(c_P+p_P+2s_P)$ & - &-\\
&$\f\o K^0$      & $p_V+s_P$ & - &-\\
\hline\hline
\end{tabular}
\caption{Same as Table~\ref{PPamplitude0} but for strangeness-conserving $B \to VP$ decays.}
\label{PVamplitude0}
\end{table}
\endgroup

\begingroup
\squeezetable
\begin{table}[t!h]
\begin{tabular}{ccccc}
\hline\hline
\multicolumn{2}{c}{Mode} & Flavor Amplitude & ~~~$BF$~~~ & ~~~$A_{CP}$~~~   \\
\hline
$B^+\to$  &$K^{*0}\p^+$     & $p_P'$                                                                                                                     & $10.1\pm0.9$ (1.28) \cite{etaKexp}     &$-0.15\pm0.07$ \cite{etaKexp}\\
&$K^{*+}\p^0$     & $-\frac{1}{\sqrt2}(t_P'+c_V'+p_P')$                                                                                          & $9.2\pm1.5$ \cite{etaKexp}                 &$-0.52\pm0.15$ \cite{etaKexp}\\
&$\r^0K^+$          & $-\frac{1}{\sqrt2}(t_V'+c_P'+p_V')$                                                                                          & $3.81^{+0.48}_{-0.46}$&$0.37\pm0.11$\\
&$\r^+K^0$          & $p_V'$                                                                                                                                      &$9.4\pm3.2$ \cite{etaKexp}      &$0.21\pm0.36$ \cite{etaKexp}\\
&$K^{*+}\e$         & $\frac{c_\phi}{\sqrt2}[t_P'+c_V'+p_P'-\sqrt2t_\phi p_V'+(-\sqrt2t_\phi+2)s_V']$                      & $19.3\pm1.6$               &$0.02\pm0.06$\\
&$K^{*+}\e'$        & $\frac{s_\phi}{\sqrt2}[t_P'+c_V'+p_P'+\frac{\sqrt2}{t_\phi}p_V'+(\frac{\sqrt2}{t_\phi}+2)s_V']$ & $5.0^{+1.8}_{-1.6}$      &$-0.26\pm0.27$\\
&$\w K^+$           & $\frac{1}{\sqrt2}(t_V'+c_P'+p_V'+2s_P')$                                                                                 & $6.5\pm0.4$ (1.11) \cite{omegaK}      &$-0.02\pm0.04$ \cite{omegaK}\\
&$\f K^+$            & $p_P'+s_P'$                                                                                                                              & $8.8\pm0.5$  (1.15)     &$0.04\pm0.02$ (1.26) \cite{phiKphipi}\\
\hline
$B^0\to$ &$K^{*+}\p^-$      & $-(t_P'+p_P')$                                                                                                            & $8.5\pm0.7$                &$-0.23\pm0.06$\\
&$K^{*0}\p^0$     & $\frac{1}{\sqrt2}(c_V'-p_P')$                                                                                                      & $2.5\pm0.6^*$ (2.52)  &$-0.15\pm0.13^*$\\
&$\r^-K^+$          & $-(t_V'+p_V')$                                                                                                                            & $7.2\pm0.9$ (1.63)      &$0.20\pm0.11$\\
&$\r^0K^0$         & $-\frac{1}{\sqrt2}(c_P'-p_V')$                                                                                                      & $4.7\pm0.7$                &$0.06\pm0.20$\\
     &                    &                                                                                                                                                    &                                     &$0.54^{+0.18}_{-0.21}$\\
&$K^{*0}\e$        & $\frac{c_\phi}{\sqrt2}[c_V'+p_P'-\sqrt2t_\phi p_V'+(-\sqrt2t_\phi+2)s_V']$                                & $15.9\pm1.0$               &$0.19\pm0.05$\\
&$K^{*0}\e'$       & $\frac{s_\phi}{\sqrt2}[c_V'+p_P'+\frac{\sqrt2}{t_\phi}p_V'+(\frac{\sqrt2}{t_\phi}+2)s_V']$           & $2.8\pm0.6$ \cite{pi0pi0exp} &$-0.07\pm0.18$ \cite{pi0pi0exp}\\
&$\w K^0$          & $\frac{1}{\sqrt2}(c_P'+p_V'+2s_P')$                                                                                           & $4.8\pm0.4$  \cite{omegaK}   & $0.04\pm0.14$ (3.04) \\
     &                   &                                                                                                                                                     &                                      & $0.71\pm0.21$ \\
&$\f K^0$           & $p_P'+s_P'$                                                                                                                                & $7.3^{+0.7}_{-0.6}$      &$-0.01\pm0.14$\\
     &                   &                                                                                                                                                     &                                      &$0.74^{+0.11}_{-0.13}$ (1.04)\\
\hline
$B^0_s\to$ &$K^{*+}K^-$      & $-(t_P'+p_P'+e_V')$  & &-\\
&$K^{*-}K^+$      & $-(t_V'+p_V'+e_P')$  &   &-\\
&\qquad$K^{*\pm}K^\mp$      & 
& $12.7\pm2.7^*$ \cite{BsVPexp} &-\\
&$K^{*0}\o K^0$ & $p_P'$  & - &-\\
&$\o K^{*0}K^0$ & $p_V'$  & - &-\\
&$\r^0\e$            & $\frac{s_\phi}{\sqrt2}c_P'-\frac{c_\phi}{2}(e_P'+e_V')$    &-  &-\\
&$\r^0\e'$           & $-\frac{c_\phi}{\sqrt2}c_P'-\frac{s_\phi}{2}(e_P'+e_V')$    & - &-\\
&$\w\e$              & $-\frac{s_\phi}{\sqrt2}(c_P'+2s_P')+\frac{c_\phi}{2}(e_P'+e_V')$ & -  &-\\
&$\w\e'$             & $\frac{c_\phi}{\sqrt2}(c_P'+2s_P')+\frac{s_\phi}{2}(e_P'+e_V')$ & - &-\\
&$\f\p^0$           & $-\frac{1}{\sqrt2}c_V'$              & - &-\\
&$\f\e$  & $-\frac{c_\phi}{\sqrt2}[-c_V'+\sqrt2t_\phi p_P'+\sqrt2t_\phi p_V'+\sqrt2t_\phi s_P'+(\sqrt2t_\phi-2)s_V']$  & - &-\\
&$\f\e'$ & $\frac{s_\phi}{\sqrt2}[c_V'+\frac{\sqrt2}{t_\phi}p_P'+\frac{\sqrt2}{t_\phi}p_V'
+\frac{\sqrt2}{t_\phi}s_P'+(\frac{\sqrt2}{t_\phi}+2)s_V']$    & - &-\\
&$\rho^+\pi^-$   &$-e_V'$ &-  &-\\
&$\rho^-\pi^+$   &$-e_P'$ &-  &-\\
&$\rho^0\pi^0$   &$\frac{1}{2}(e_P'+e_V')$ &-  &-\\
&$\omega\pi^0$   &$-\frac{1}{2}(e_P'+e_V')$ &-  &-\\
\hline\hline
\end{tabular}
\caption{Same as Table~\ref{PPamplitude0} but for strangeness-changing $B \to VP$ decays.}
\label{PVamplitude1}
\end{table}
\endgroup

Current experimental data on branching fractions and CP asymmetries as well as the flavor amplitude decomposition for all the $B \to VP$ decays are given in Table~\ref{PVamplitude0} and Table~\ref{PVamplitude1} for strangeness-conserving and strangeness-changing transitions, respectively.  There are totally 23 theory parameters to fit in this sector, 15 of them are involved in Scheme A.  We perform three types of fits here.  Scheme A is limited to those modes not involving the flavor-singlet amplitudes.  Scheme B is a global fit to all $VP$ data points using all the theory parameters except for the $W$-exchange amplitudes.  Finally, the $E_{P,V}$ amplitudes are included in the global fit of Scheme C.

We first enumerate the data points not included in our $\chi^2$ fits.  There is a large scale factor in the branching fraction of $B^0\to K^{*0}\pi^0$ and its CP asymmetry is only reported by BaBar \cite{Kstarpi}.  As we will see, all theoretical calculations predict a negative CP asymmetry for the $B^0\to\rho^0\pi^+$ decay, whereas the data, only measured by BaBar \cite{rho0pip}, give the opposite sign.  Therefore, we remove these data points from the fits to improve the fit quality.  We then have 27 observables for 15 parameters in Scheme A, 51 observables for 19 parameters in Scheme B, and 51 observables for 23 parameters in Scheme C.  As in the $PP$ sector, all the SU(3) breaking factors $\xi$'s are fixed at unity in all the presented schemes.

\begingroup
\begin{table}[h!t]
\begin{tabular}{ccccc}
\hline\hline
 Observable   & ~~~~~~BaBar~~~~~~ & ~~~~~~Belle~~~~~~& ~~~~~~CLEO~~~~~~ & ~~~~~~Average~~~~~~ \\
\hline\hline
$BF^{\pm\mp}_{\r\pi}$ & $22.6\pm1.8\pm2.2$ & $22.6\pm1.1\pm4.4$
& $27.6^{+8.4}_{-7.4}\pm4.2$ & $23.0\pm2.3$ \\
$A_{\r\pi}$   &$-0.10\pm0.03\pm0.02$ &$-0.12\pm0.05\pm0.04$ &&$-0.11\pm0.03$\\
$C$            &$0.02\pm0.06\pm0.04$ &$-0.13\pm0.09\pm0.05$ &&$-0.03\pm0.06$\\
$S$            &$0.05\pm0.08\pm0.03$ &$0.06\pm0.13\pm0.05$ &&$0.06\pm0.07$\\
$\Delta C$ &$0.23\pm0.06\pm0.05$ &$0.36\pm0.10\pm0.05$   &&$0.27\pm0.06$\\
$\Delta S$ &$0.05\pm0.08\pm0.04$ &$-0.08\pm0.13\pm0.05$ &&$0.01\pm0.08$\\
\hline\hline
\end{tabular}
\caption{Branching fractions and time-dependent CP asymmetries of the $B^0\to\rho^\pm\pi^\mp$ decays.}
\label{rhopiexp}
\end{table}
\endgroup

Among all the data points, some conversion has to be done for the $B^0 \to \rho^\pm \pi^\mp$ observables as experimental data do not directly provide the quantities required by us.  First, we extract individual branching fractions of $B^0\to\rho^-\pi^+$ and $B^0\to\rho^+\pi^-$ according to
\begin{equation}
\begin{split}
BF(B^0\to\rho^-\pi^+) &=\frac{1}{2}(1-\Delta C-A_{\r\pi}C)BF^{\pm\mp}_{\r\pi}=8.4\pm1.1~,\\
BF(B^0\to\rho^+\pi^-) &=\frac{1}{2}(1+\Delta C+A_{\r\pi}C)BF^{\pm\mp}_{\r\pi}=4.6\pm1.6 ~,
\end{split}
\label{rhopiBr}
\end{equation}
where the experimental data for $BF^{\pm\mp}_{\rho\pi}$, $C$, $\Delta C$ and $A_{\rho\pi}$ are given in Table~\ref{rhopiexp}.
Time-dependent CP asymmetries are given by
\begin{equation}
\begin{split}
{\cal A}_{+-}
&\equiv {\cal A}(B^0\to\r^+\pi^-)
= -\frac{A_{\r\pi}+C+A_{\r\pi}\Delta C}{1+\Delta C+A_{\r\pi}C}=0.13\pm0.06 \,,\\
{\cal A}_{-+}
&\equiv {\cal A}(B^0\to\r^-\pi^+)
=\frac{A_{\r\pi}-C-A_{\r\pi}\Delta C}{1-\Delta C-A_{\r\pi}C}=-0.07\pm0.09 \,,
\end{split}
\label{rhopiAcp}
\end{equation}
and
\begin{equation}
\begin{split}
{\cal S}_{+-} &\equiv {\cal S}(B^0\to\rho^+\pi^-) = S + \Delta S ~, \\
{\cal S}_{-+} &\equiv {\cal S}(B^0\to\rho^-\pi^+) = S - \Delta S ~,
\end{split}
\label{Spm}
\end{equation}
where $S$ and $\Delta S$ can also be found in Table~\ref{rhopiexp}.
Note that theoretical values of the mixing-induced CP asymmetries are
\begin{equation}
\begin{split}
{\cal S}_{+-} &=\frac{2\text{Im}[\lambda^{+-}]}{1+|\lambda^{+-}|^2}\hspace{5pt},\hspace{5pt}\lambda^{+-}=\frac{V_{td}}{V^*_{td}}\frac{|\o{t}_V+\o{p}_V|}{|t_P+p_P|} ~, \\
{\cal S}_{-+} &=\frac{2\text{Im}[\lambda^{-+}]}{1+|\lambda^{-+}|^2}\hspace{5pt},\hspace{5pt}\lambda^{-+}=\frac{V_{td}}{V^*_{td}}\frac{|\o{t}_P+\o{p}_P|}{|t_V+p_V|} ~,
\end{split}
\label{rhopiS}
\end{equation}
where a bar over the amplitudes denotes CP conjugation.

\subsection{Fit Results \label{sec:PVscheme}}

\begingroup
\begin{table}[t!h]
\begin{tabular}{cccc}
\hline\hline
 ~~Parameter~~  & ~~Scheme A ~~       & ~~
Scheme B~~ &
 ~~Scheme C~~  \\
\hline
   $|T_P|$ & $1.173^{+0.063}_{-0.066}$ & $1.193^{+0.060}_{-0.063}$  &$0.909^{+0.499}_{-0.331}$  \\
   $|T_V|$ & $0.880^{+0.058}_{-0.063}$ & $0.883^{+0.057}_{-0.060}$  &$0.704^{+0.294}_{-0.275}$  \\
   $\delta_{T_V}$ & $3\pm4$                  & $1\pm4$                                &$-6^{+28}_{-39}$ \\
   $|C_P|$ & $0.341^{+0.135}_{-0.130}$ & $0.284^{+0.092}_{-0.081}$  &$0.524^{+0.294}_{-0.301}$  \\
   $\delta_{C_P}$ & $-24^{+41}_{-32}$   & $-36^{+29}_{-23}$                &$-54^{+32}_{-44}$  \\
   $|C_V|$ & $0.668^{+0.325}_{-0.276}$ & $0.735^{+0.164}_{-0.161}$ & $1.120^{+0.416}_{-0.339}$ \\
   $\delta_{C_V}$ & $-89^{+27}_{-16}$   & $-91^{+13}_{-10}$                &$-93^{+15}_{-17}$  \\
   $|P_P|$ & $0.083\pm0.003$                &  $0.083\pm0.002$                &$0.083\pm0.003$ \\
   $\delta_{P_P}$ & $-25\pm6$               &  $-21\pm5$                            &$-37^{+17}_{-39}$   \\
   $|P_V|$            & $0.066\pm0.005$     &  $0.069\pm0.004$                 &$0.070\pm0.004$ \\
   $\delta_{P_V}$ & $165\pm9$               & $159^{+7}_{-8}$                    &$142^{+17}_{-35}$ \\
   $|P_{EW,P}|$    & $0.035^{+0.010}_{-0.011}$  & $0.031\pm0.010$   &$0.030^{+0.009}_{-0.010}$ \\
   $\delta_{PEW,P}$ & $51^{+12}_{-16}$          & $44^{+11}_{-15}$                   &$25^{+20}_{-35}$ \\
   $|P_{EW,V}|$ & $0.061^{+0.029}_{-0.024}$  &$0.058^{+0.017}_{-0.015}$    &$0.064^{+0.020}_{-0.018}$ \\
   $\delta_{PEW,V}$ & $-100^{+35}_{-23}$       & $-83^{+22}_{-15}$                 &$-105^{+26}_{-34}$ \\
   $|S_P|$            & -                                          & $0.015^{+0.006}_{-0.005}$   &$0.014\pm0.006$  \\
   $\delta_{S_P}$ & -                                          &$-142^{+13}_{-21}$               &$-154^{+21}_{-38}$  \\
   $|S_V|$            & -                                           &$0.033\pm0.004$                &$0.035^{+0.005}_{-0.004}$  \\
   $\delta_{S_V}$ & -                                           &$-73\pm24$                         &$-89^{+24}_{-27}$   \\
   $|E_P|$             & -                                          &-                                           & $0.266^{+0.829}_{-0.266}$ \\
   $\delta_{E_P}$ & -                                           &-                                           & $120\pm180$ \\
   $|E_V|$             & -                                           & -                                         &$0.467^{+0.526}_{-0.375}$ \\
   $\delta_{E_V}$ & -                                           &-                                           &$-65^{+27}_{-86}$     \\
\hline
   $\chi^2_{min}/dof$&15.53/12      &40.22/32       &37.57/28  \\
    Fit quality &12.36 \%      &15.08\%          &10.67\%  \\
\hline\hline
\end{tabular}
\caption{Fit results of theory parameters.  Different fit schemes are defined in the text.  Magnitudes of the amplitudes are quoted in units of $10^4$ eV, and the strong phases in units of degree.}
\label{PVall}
\end{table}
\endgroup

Table~\ref{PVall} summarizes the results of the three fits.  This sector also shows a general hierarchy: $|T_{P,V}| > |C_{P,V}| > |P_{P,V}| > |P_{EW,P}|~,~ |S_{P,V}|$ and yet $|P_{EW,V}|\sim |P_V|$.  The fit quality drops as we include the $S_{P,V}$ and $E_{P,V}$ amplitudes.  The uncertainties in many parameters in Scheme~C, notably the color-allowed and color-suppressed tree amplitudes, are significantly larger than the corresponding ones in the other two schemes.  Moreover, the large error bars on the magnitudes and phases of the $E_{P,V}$ amplitudes in Scheme~C suggest that the current data precision is unable to fix these amplitudes well.  Therefore, we consider the fit in Scheme~C less reliable, and will instead use Scheme B as the preferred one in our later discussions and predictions.  We have tried and observed that there is no significant improvement in fit quality by including the SU(3) breaking factors, which are found to be consistent with unity.  We have also tried a global fit as in Scheme B but with the mixing angle $\phi$ free to vary.  It is found that $\phi=(43\pm6)^\circ$, and the fit quality decreases slightly to $13\%$.

The QCD penguin amplitudes are quite stable across the fits, with the penguin-dominated $B^+\to K^{*0}\pi^+$ and $B^+\to\rho^+K^0$ decay being essential in fixing $|P_P|$ and $|P_V|$, respectively.  The relative strong phase between $T_V$ and $T_P$ is consistent with zero.  Doing a global fit without the flavor-singlet amplitudes gives essentially the same values for most parameters as in Scheme~A, except for $C_V$ and $P_{EW,V}$, but has a much worse fit quality, indicating a strong need for the $S_P$ and $S_V$ amplitudes.  Unlike what we obtain in the $PP$ sector, the magnitude of $S_{P,V}$ are smaller than $P_{EW,P,V}$, as shown in Scheme~B.

The major differences between the restricted fit in Scheme A and the global fit in Scheme B are in the color-suppressed tree and EW penguin amplitudes.  Going from Scheme~A to Scheme~B, the central value of $|C_{P}|$ reduces slightly while that of $|C_V|$ increases.  Correspondingly, $P_{EW,P}$ and $P_{EW,V}$ also have changes in both sizes and phases.  It is noticed that the error bars associated with $|C_{P,V}|$ and $|P_{EW,PV}|$ are the largest among all parameters, about $25\%$ to $30\%$.  An immediate consequence of such large uncertainties is that our predictions for modes involving these amplitudes tend to have larger errors {\it e.g.}, $BF(B^0 \to \rho^0 \pi^0)$, $BF(B^+ \to K^{*+} \pi^0)$ and $BF(B^+ \to \rho^0 K^+)$.

\begingroup
\begin{table}[h!t]
\begin{tabular}{cccc}
\hline\hline
        & ~~~~~~Scheme A~~~~~~ & ~~~~~~Scheme B~~~~~~ \\
\hline
   $|C_V/T_P|$ &$0.57\pm0.26$ & $0.62\pm0.14$ \\
   $|C_P/T_V|$ & $0.39\pm0.15$ &$0.32\pm0.10$\\
\hline\hline
\end{tabular}
\caption{Magnitudes of the ratios of color-suppressed tree amplitude to the color-allowed tree amplitude based on different schemes in Table~\ref{PVall}.}
\label{CTVP}
\end{table}
\endgroup

As the $C$ amplitude in the $PP$ sector, the $C_V$ amplitude has a large size and is about twice larger in magnitude than the $C_P$ amplitude.  The ratios of $|C_V/T_P|$ and $|C_P/T_V|$ in Schemes A and B are given in Table~\ref{CTVP}.  Although with large errors, $C_P$ and $C_V$ have strong phases around $-30^\circ$ and $-90^\circ$, respectively, relative to the color-allowed tree amplitudes.  Also related to the fact of $|C_V/C_P| \simeq 2$, our fit results show that $|P_{EW,V} / P_{EW,P}|~,~ |S_V / S_P| \simeq 2$ as well.  A further comparison of the color-suppressed tree amplitudes to the color-allowed tree amplitudes will be made in Section~\ref{sec:PPQCDFsec}.

The QCD penguin amplitudes are about one order of magnitude smaller than the color-allowed tree amplitudes, with $|P_P|$ slightly larger than $|P_V|$.  It is noted that $P_P$ and $P_V$ are almost opposite in phase, in agreement with the proposal made in Ref.~\cite{Lipkin}.  This property results in constructive and destructive interference effects in the $\eta K^*$ and $\eta' K^*$ modes, respectively.  Similar effects on several $\Delta S = 0$ decays ({\it e.g.}, $\rho^{0,0,+}\pi^{0,+,0}$, $\eta\rho$, $\eta'\rho$, $\cdots$ etc.) are less prominent because of the CKM factor suppression.  Besides, $P_P$ has only a small strong phase of $\sim -20^\circ$ relative to $T_P$, so that $P_V$ is almost opposite to both $T_P$ and $T_V$.  This leads to a significant interference effect on modes involving the color-allowed tree and QCD penguin amplitudes.  For example, as given in the next subsection, the $B_s \to \rho^+ K^-$ is predicted to have the largest branching fraction of order $15\times 10^{-6}$ among the $B_s\to VP$ decays.

One of the striking features in our diagrammatic analysis is that the electroweak penguin amplitude $P_{EW,V}$ is comparable in magnitude to the QCD penguin amplitude $P_V$ (see Table~\ref{PVall}).  In contrast, $|P_{EW}|$ is suppressed by one order of magnitude relative to $|P|$ in the $PP$ sector.  This observation has some important implications for CP violation in the $K^*\pi$ modes and for the branching fractions of $B_s\to\phi\pi^0$ (and $\phi\rho^0$ as well).  In the absence of the $c'_V$ amplitude, we see from Table~\ref{PVamplitude1} that the $K^{*+}\pi^0$ and $K^{*+}\pi^-$ decays should have the same CP asymmetry. Just as in the $B\to K\pi$ decays, a sign flip in $A_{CP}(K^{*+}\pi^0)$ will occur in the presence of a large complex $C_V$ (this can be checked by using any value of $C_V$ extracted in Table~\ref{PVall}).  This is in contradiction with the experimental observation that CP asymmetries of $K^{*+}\pi^0$ and $K^{*+}\pi^-$ are of the same sign.  This enigma can be solved by noting that $c'_V=Y_{sb}^u C_V-(Y_{sb}^u+Y_{sb}^c)P_{EW,V}$.  Since $|Y_{sb}^c| \gg |Y_{sb}^u|$ and $|P_{EW,V}| \sim |P_V|$, the $P_{EW,V}$ amplitude will make a substantial contribution to $c'_V$ and render $A_{CP}(K^{*+}\pi^0)$ a correct sign.  In the $K\pi$ case, the electroweak penguin $P_{EW}$ is suppressed relative to the QCD penguin $P$.  It will affect the magnitude of $A_{CP}(K^+\pi^0)$, but not its sign.

\subsection{Predictions \label{sec:prediction}}

\begingroup
\squeezetable
\begin{table}[t!p]
\begin{tabular}{ccccccc}
\hline\hline
\multicolumn{2}{c}{Mode}   & Data   & This Work      & QCDF & pQCD   &SCET            \\
         \hline
  $B^{+,0}\to$  &$\o K^{*0}K^+$ &$<1.1$ &$0.46\pm0.03$  &$0.80^{+0.20+0.31}_{-0.17-0.28}$           &$0.32^{+0.12}_{-0.08}$ \cite{KKPVp}                                                    &$0.51^{+0.18+0.07}_{-0.16-0.06}$\\
  &$K^{*+}\o K^0$&-                                 &$0.31\pm0.03$  &$0.46^{+0.37+0.42}_{-0.17-0.26}$             &$0.21^{+0.14}_{-0.12}$ \cite{KKPVp}                                                  &$0.51^{+0.21+0.08}_{-0.17-0.07}$\\
  &$\r^0\p^+$       &$8.3^{+1.2}_{-1.3}$    &$7.59\pm1.41$  &$8.7^{+2.7+1.7}_{-1.3-1.4}$                       &$\sim 9.3$ \cite{Li:2014haa}                                             &$7.9^{+0.2}_{-0.1}\pm0.8$\\
  &$\r^+\p^0$       &$10.9^{+1.4}_{-1.5}$  &$12.15\pm2.52$&$11.8^{+1.8}_{-1.1}\pm1.4$                        &$\sim 7.2$ \cite{Li:2014haa}                                               &$11.4\pm0.6^{+1.1}_{-0.9}$\\
  &$\r^+\e$           &$6.9\pm1.0$               &$5.26\pm1.19$  &$8.3^{+1.0}_{-0.6}\pm0.9$                         &$6.7^{+2.6}_{-1.9}$ \cite{etaPVp}                                                         &$3.3^{+1.9}_{-1.6}\pm0.3$\\
  &$\r^+\e'$          &$9.8^{+2.1}_{-2.0}$     &$5.66\pm1.25$  &$5.6^{+0.9+0.8}_{-0.5-0.7}$                      &$4.6^{+1.6}_{-1.4}$ \cite{etaPVp}                                                         &$0.44^{+3.18+0.06}_{-0.20-0.05}$\\
  &$\w\p^+$         &$6.9\pm0.5$                &$7.03\pm1.42$  &$6.7^{+2.1+1.3}_{-1.0-1.1}$                      &$\sim 6.1$  \cite{Li:2014haa}                                                &$8.5\pm0.3\pm0.8$\\
  &$\f\p^+$           &$<0.15$                       &$0.04\pm0.02$  &$\approx0.043$                                         &$0.032^{+0.008-0.012}_{-0.007+0.018}$ \cite{phipiPVp}                     &$\approx0.003$\\
  &$\o K^{*0}K^0$&-                                   &$0.43\pm0.02$  &$0.70^{+0.18+0.28}_{-0.15-0.25}$           &$0.24\pm0.02^{+0.00+0.03+0.06}_{-0.01-0.04-0.04}$ \cite{KKPVp}    &$0.47^{+0.17+0.06}_{-0.14-0.05}$\\
 &$K^{*0}\o K^0$&$<1.9$                         &$0.29\pm0.03$  &$0.47^{+0.36+0.43}_{-0.17-0.27}$           &$0.49^{+0.12+0.03+0.05+0.03}_{-0.08-0.02-0.04-0.01}$ \cite{KKPVp}&$0.48^{+0.20+0.07}_{-0.16-0.06}$\\
 &$\r^-\p^+$        &$8.4\pm1.1$                 &$8.22\pm1.06$  &$9.2^{+0.4+0.5}_{-0.7-0.7}$                     &$\sim 10.7$ \cite{Li:2014haa}                                               &$6.6^{+0.2}_{-0.1}\pm0.7$\\
 &$\r^+\p^-$        &$14.6\pm1.6$               &$15.20\pm1.52$&$15.9^{+1.1+0.9}_{-1.5-1.1}$                   &$\sim 20.1$ \cite{Li:2014haa}                                                &$10.2^{+0.4}_{-0.5}\pm0.9$\\
 &$\r^0\p^0$       &$2.0\pm0.5$                  &$2.24\pm0.93$ &$1.3^{+1.7+1.2}_{-0.6-0.6}$                     &$\sim 1.1$ \cite{Li:2014haa}                                               &$1.5\pm0.1\pm0.1$\\
 &$\r^0\e$           &$<1.5$                           &$0.54\pm0.32$ &$0.10^{+0.02+0.04}_{-0.01-0.03}$           &$0.13^{+0.13}_{-0.06}$ \cite{etaPVp}                                                   &$0.14^{+0.33}_{-0.13}\pm0.01$\\
 &$\r^0\e'$          &$<1.3$                           &$0.63\pm0.33$ &$0.09^{+0.10+0.07}_{-0.04-0.03}$           &$0.10\pm0.05$ \cite{etaPVp}                                                                &$1.0^{+3.5}_{-0.9}\pm0.1$\\
 &$\w\p^0$         &$<0.5$                           &$1.02\pm0.66$ &$0.01^{+0.02+0.04}_{-0.00-0.01}$           & $\sim 0.85$ \cite{Li:2014haa}                                                                                         &$0.015^{+0.024}_{-0.000}\pm0.002$\\
 &$\w\e$             &$<1.4$                           &$1.12\pm0.44$ &$0.85^{+0.65+0.40}_{-0.26-0.24}$          &$0.71^{+0.37}_{-0.28}$ \cite{etaPVp}                                                   &$1.4^{+0.8}_{-0.6}\pm0.1$\\
 &$\w\e'$            &$<1.8$                           &$1.24\pm0.47$ &$0.59^{+0.50+0.33}_{-0.20-0.18}$          &$0.55^{+0.31}_{-0.26}$ \cite{etaPVp}                                                   &$3.1^{+4.9}_{-2.6}\pm0.3$\\
 &$\f\p^0$           &$<0.15$                         &$0.02\pm0.01$ &$0.01^{+0.03+0.02}_{-0.01-0.01}$          &$0.0068\pm0.0003^{-0.0007}_{+0.0010}$ \cite{phipiPVp}                   &$\approx0.001$\\
 &$\f\e$               &$<0.5$                           &$0.01\pm0.01$&$\approx0.005$                                        &$0.011^{+0.062}_{-0.009}$ \cite{etaPVp}                                             &$\approx0.0008$\\
&$\f\e'$               &$<0.5$                          &$0.01\pm0.01$&$\approx0.004$                                         &$0.017^{+0.161}_{-0.010}$ \cite{etaPVp}                                             &$\approx0.0007$\\
 &$K^{*0}\p^+$   &$10.1\pm0.9$               &$10.47\pm0.60$&$10.4^{+1.3+4.3}_{-1.5-3.9}$                  &$6.0^{+2.8}_{-1.5}$ \cite{PVp}                                                              &$9.9^{+3.5+1.3}_{-3.0-1.1}$\\
 &$K^{*+}\p^0$   &$9.2\pm1.5$                  &$9.79\pm2.95$  &$6.7\pm0.7^{+2.4}_{-2.2}$                    &$4.3^{+5.0}_{-2.2}$ \cite{PVp}                                                               &$6.5^{+1.9}_{-1.7}\pm0.7$\\
 &$\r^0K^+$        &$3.81^{+0.48}_{-0.46}$&$3.97\pm0.90$  &$3.5^{+2.9+2.9}_{-1.2-1.8}$                  &$5.1^{+4.1}_{-2.8}$ \cite{PVp}                                                               &$4.6^{+1.8+0.7}_{-1.5-0.6}$\\
 &$\r^+K^0$        &$9.4\pm3.2$                 &$7.09\pm0.77$  &$7.8^{+6.3+7.3}_{-2.9-4.4}$  &$8.7^{+6.8}_{-4.4}$ \cite{PVp}                                                               &$10.1^{+4.0+1.5}_{-3.3-1.3}$\\
 &$K^{*+}\e$       &$19.3\pm1.6$               &$16.57\pm2.58$&$15.8^{+8.2+9.6}_{-4.2-7.3}$ \cite{etaQ}&$22.13^{+0.26}_{-0.27}$ \cite{etaKp}                                                     &$18.6^{+4.5+2.5}_{-4.8-2.2}$\\
 &$K^{*+}\e'$      &$5.0^{+1.8}_{-1.6}$       &$3.43\pm1.43$  &$1.6^{+2.1+3.7}_{-0.3-1.6}$ \cite{etaQ} &$6.38\pm0.26$ \cite{etaKp}                                                                    &$4.8^{+5.3+0.8}_{-3.7-0.6}$\\
 &$\w K^+$          &$6.5\pm0.4$                  &$6.43\pm1.49$  &$4.8^{+4.4+3.5}_{-1.9-2.3}$                  &$10.6^{+10.4}_{-5.8}$ \cite{PVp}                                                            &$5.9^{+2.1+0.8}_{-1.7-0.7}$\\
 &$\f K^+$            &$8.8\pm0.5$                  &$8.34\pm1.31$  &$8.8^{+2.8+4.7}_{-2.7-3.6}$                  &$7.8^{+5.9}_{-1.8}$ \cite{PVp}                                                                &$8.6^{+3.2+1.2}_{-2.7-1.0}$\\
 &$K^{*+}\p^-$     &$8.5\pm0.7$                  &$8.35\pm0.50$  &$9.2\pm1.0^{+3.7}_{-3.3}$                     &$6.0^{+6.8}_{-2.6}$ \cite{PVp}                                                                &$9.5^{+3.2+1.2}_{-2.8-1.1}$\\
 &$K^{*0}\p^0$    &$2.5\pm0.6$                  &$3.89\pm1.98$  &$3.5\pm0.4^{+1.6}_{-1.4}$                     &$2.0^{+1.2}_{-0.6}$ \cite{PVp}                                                                &$3.7^{+1.4}_{-1.2}\pm0.5$\\
 &$\r^-K^+$          &$7.2\pm0.9$                 &$8.28\pm0.80$  &$8.6^{+5.7+7.4}_{-2.8-4.5}$                   &$8.8^{+6.8}_{-4.5}$ \cite{PVp}                                                                &$10.2^{+3.8+1.5}_{-3.2-1.2}$\\
 &$\r^0K^0$         &$4.7\pm0.7$                 &$4.97\pm1.14$  &$5.4^{+3.4+4.3}_{-1.7-2.8}$                    &$4.8^{+4.3}_{-2.3}$ \cite{PVp}                                                               &$5.8^{+2.1+0.8}_{-1.8-0.7}$\\
 &$K^{*0}\e$       &$15.9\pm1.0$               &$16.34\pm2.48$&$15.7^{+7.7+9.6}_{-4.0-7.3}$ \cite{etaQ}&$22.31^{+0.28}_{-0.29}$ \cite{etaKp}                                                    &$16.5^{+4.1+2.3}_{-4.3-2.0}$\\
 &$K^{*0}\e'$      &$2.8\pm0.6$                  &$3.14\pm1.24$  &$1.5^{+1.8+3.5}_{-0.3-1.6}$ \cite{etaQ}   &$3.35^{+0.29}_{-0.27}$ \cite{etaKp}                                                     &$4.0^{+4.7+0.7}_{-3.4-0.6}$\\
 &$\w K^0$          &$4.8\pm0.4$                 &$4.82\pm1.26$  &$4.1^{+4.2+3.3}_{-1.7-2.2}$                     &$9.8^{+8.6}_{-4.9}$ \cite{PVp}                                                             &$4.9^{+1.9+0.7}_{-1.6-0.6}$\\
 &$\f K^0$           &$7.3^{+0.7}_{-0.6}$       &$7.72\pm1.21$  &$8.1^{+2.6+4.4}_{-2.5-3.3}$                     &$7.3^{+5.4}_{-1.6}$ \cite{PVp}                                                            &$8.0^{+3.0+1.1}_{-2.5-1.0}$\\
\hline\hline
\end{tabular}
\caption{Predicted branching fractions (in units of $\times10^{-6}$) of all the $B^{+,0}$ decays using the fit results of Scheme B.  All the predictions made by QCDF and SCET are taken from Ref.~\cite{Q} and work 2 of Ref.~\cite{PVS}, respectively.  The pQCD predictions taken from \cite{Li:2014haa} are for $S_e=-\pi/2$ with $S_e$ being a strong phase induced by Glauber gluons.  We have followed the prescription outlined in Sec.~V to convert the $B^0\to\rho^\pm\pi^\mp$ observables in Ref.~\cite{Li:2014haa} into the ones for $B^0\to \rho^+\pi^-$ and $B^0\to\rho^-\pi^+$. }
\label{preglobalBr}
\end{table}
\endgroup

Tables~\ref{preglobalBr} to \ref{preglobalsAcp} present our predictions for all the $B\to VP$ observables based on Scheme~B in Table~\ref{PVall}, along with those made in the QCDF, pQCD and SCET approaches.  In the following, we highlight observables where there exist disparities among our predictions, experimental data and other theoretical calculations.

Table~\ref{preglobalBr} shows the branching fractions of all the $B^{+,0}$ decays.  Compared to the current data, all theoretical calculations including ours expect a smaller branching fraction for $B^+ \to\rho^+\eta'$.  Since the penguin amplitudes are much less important, the $B^+ \to\rho^+\eta'$ decay rate should be about half that of $\rho^+ \pi^0$  and about the same as that of $\rho^+ \eta$ with our choice of $\phi$.

Since the $B^0\to\rho^0\pi^0$, $\rho^0 \eta^{(\prime)}$, $\omega\pi^0$ and $\omega\eta^{(\prime)}$ decays are dominated by the color-suppressed tree amplitudes, their branching fractions obtained in the perturbative approaches are generally smaller than ours.  Furthermore, our prediction of $BF(B^0\to\rho^0\pi^0)$ agrees well with the measured value.  This is mainly the result of partially constructive interference between the $C_P$ and $C_V$ amplitudes that subtend a relative phase less than $90^\circ$.  It is noted that our prediction of $BF(B^0\to\omega\pi^0)$ is about twice larger than the current 90\% CL upper bound.  Also noted is that $BF(B^0\to\omega\pi^0) \simeq 2 BF(B^0\to\rho^0 \eta)~,~2 BF(B^0\to\rho^0 \eta')$ when the QCD penguin amplitudes are neglected.  By a similar token, $BF(B^0 \to \omega\eta)$ and $BF(B^0 \to \omega\eta')$ are about half $BF(B^0 \to \rho^0\pi^0)$ in our work.  Yet perturbative calculations have more diverse predictions on $BF(B^0 \to \omega\eta')$.

Na{\"i}vely, it is expected that $B^0\to K^{*+}\pi^-$ has a rate larger than that of $B^+\to K^{*+}\pi^0$ owing to the wavefunction of the $\pi^0$.  Indeed, this is the pattern predicted by all the existing perturbative approaches in Table~\ref{preglobalBr}.
However, the experimental measurements and our fit results indicate that their rates are comparable and the latter has even a slightly larger branching fraction.  This has to do with the sizeable $c'_V$ amplitude which contributes constructively to $B^+\to K^{*+}\pi^0$.
The $B^+\to K^{*0}\pi^+$ decay involves purely the $p'_P$ amplitude.  All the theoretical calculations except the pQCD give roughly the same branching fraction as the measured value.  The branching fraction of $B^0 \to \o K^{*0} K^0$ decay can be estimated under flavor SU(3) symmetry to be about $0.43 \times 10^{-6}$, which agrees with the SCET prediction.  The $B^0 \to K^{*0} \o K^0$ and $B^+ \to \rho^+ K^0$ decays involve only the $p_V$ and $p'_V$ amplitude, respectively.  Therefore, the branching fraction of the former can be inferred by the SU(3) symmetry from that of the latter to be about $0.29 \times 10^{-6}$.  In comparison, all the perturbative calculations have a prediction of central value at about $0.5 \times 10^{-6}$.  A determination of these yet measured modes can test the SU(3) symmetry and theories.

\begingroup
\squeezetable
\begin{table}[t!]
\begin{tabular}{ccccccc}
\hline\hline
    &Mode   & Data        & This Work         & QCDF  & pQCD   & SCET             \\
\hline
$B_{+,0}\to$  &$\o K^{*0}K^+$ &-              &0                             &$-0.089\pm0.011^{+0.028}_{-0.024}$       &$-0.069^{+0.115}_{-0.103}$ \cite{KKPVp}                                                              &$-0.044\pm0.041\pm0.002$ \\
  &$K^{*+}\o K^0$&-                                   &0                             &$-0.078^{+0.059+0.041}_{-0.041-0.100}$&$0.065^{+0.123}_{-0.114}$ \cite{KKPVp}                                                                &$-0.012\pm0.017\pm0.001$\\
  &$\r^0\p^+$       &$0.18^{+0.09}_{-0.17}$&$-0.239\pm0.084$  &$-0.098^{+0.034+0.114}_{-0.026-0.102}$&$\sim -0.31$ \cite{Li:2014haa}                                                                 &$-0.192^{+0.155+0.017}_{-0.134-0.019}$\\
  &$\r^+\p^0$       &$0.02\pm0.11$             &$0.053\pm0.094$  &$0.097^{+0.021+0.080}_{-0.031-0.103}$ &$\sim 0.13$ \cite{Li:2014haa}                                                                  &$0.123^{+0.094+0.009}_{-0.100-0.011}$\\
  &$\r^+\e$           &$0.11\pm0.11$             &$0.162\pm0.072$   &$-0.085\pm0.004^{+0.065}_{-0.053}$      &$0.019^{+0.001+0.002+0.001+0.006}_{-0.000-0.003-0.000-0.005}$ \cite{etaPVp} &$-0.091^{+0.167+0.009}_{-0.158-0.008}$\\
  &$\r^+\e'$          &$0.26\pm0.17$             &$0.223\pm0.137$   &$0.014^{+0.008+0.140}_{-0.022-0.117}$ &$-0.250^{+0.004+0.041+0.008+0.021}_{-0.003-0.016-0.007-0.018}$ \cite{etaPVp}&$-0.217^{+1.359+0.021}_{-0.243-0.017}$\\
  &$\w\p^+$         &$-0.02\pm0.06$            &$0.075\pm0.067$   &$-0.132^{+0.032+0.120}_{-0.021-0.107}$&$\sim -0.18$ \cite{Li:2014haa}                                                                      &$0.023^{+0.134}_{-0.132}\pm0.002$\\
  &$\f\p^+$           &-                                    &0                             &0                                                               &$-0.080^{+0.009+0.015}_{-0.010-0.001}$ \cite{phipiPVp}                                        &\\
  &$\o K^{*0}K^0$&-                                    &0                             &$-0.135^{+0.016+0.014}_{-0.017-0.023}$&&$-0.044\pm0.041\pm0.002$\\
 &$K^{*0}\o K^0$&-                                    &0                             &$-0.035^{+0.013+0.007}_{-0.017-0.020}$&&$-0.012\pm0.017\pm0.001$\\
 &$\r^-\p^+$        &$-0.07\pm0.09$             &$-0.136\pm0.053$&$-0.227^{+0.009+0.082}_{-0.011-0.044}$&$\sim -0.27$ \cite{Li:2014haa}&$-0.124^{+0.176+0.011}_{-0.153-0.012}$\\
 &$\r^+\p^-$        &$0.13\pm0.06$              &$0.120\pm0.027$ &$0.044\pm0.003^{+0.058}_{-0.068}$        &$\sim 0.05$ \cite{Li:2014haa}&$0.108^{+0.094+0.009}_{-0.102-0.010}$\\
 &$\r^0\p^0$       &$-0.27\pm0.24$              &$-0.043\pm0.121$&$0.110^{+0.050+0.235}_{-0.057-0.288}$ &$\sim 0.18$ \cite{Li:2014haa}                                          &$-0.035^{+0.214}_{-0.203}\pm0.003$\\
 &$\r^0\e$           &-                                      &$-0.264\pm0.215$&$0.862^{+0.037+0.104}_{-0.058-0.214}$ &$-0.896^{+0.019+0.137+0.007+0.046}_{-0.009-0.039-0.001-0.090}$ \cite{etaPVp}&$0.333^{+0.669+0.031}_{-0.624-0.028}$ \\
 &$\r^0\e'$          &-                                      &$-0.440\pm0.317$&$0.535^{+0.045+0.395}_{-0.079-0.576}$ &$-0.757^{+0.056+0.131+0.063+0.129}_{-0.048-0.070-0.040-0.099}$ \cite{etaPVp}&$0.522^{+0.199+0.044}_{-0.806-0.041}$\\
 &$\w\p^0$         &-                                      &$-0.188\pm0.185$&$-0.170^{+0.554+0.986}_{-0.228-0.823}$&$\sim -0.12$ \cite{Li:2014haa}                                                                 &$0.395^{+0.791+0.034}_{-1.855-0.031}$\\
 &$\w\e$            &-                                      &$0.054\pm0.137$&$-0.447^{+0.131+0.177}_{-0.099-0.116}$ &$0.335^{+0.010+0.008+0.059+0.039}_{-0.014-0.046-0.068-0.044}$ \cite{etaPVp} &$-0.096^{+0.178}_{-0.168}\pm0.009$ \\
 &$\w\e'$           &-                                      &$-0.005\pm0.259$&$-0.414^{+0.025+0.195}_{-0.024-0.144}$ &$0.160^{+0.001+0.033+0.022+0.017}_{-0.009-0.039-0.032-0.020}$ \cite{etaPVp} &$-0.272^{+0.181+0.024}_{-0.297-0.022}$\\
 &$\f\p^0$         &-                                      &0                           &0                                                                 &$-0.063^{-0.005}_{+0.007}\pm0.025$ \cite{phipiPVp}                                              &\\
 &$\f\e$             &-                                      &0                           &0                                                                 &0 \cite{etaPVp}&\\
 &$\f\e'$            &-                                      &0                           &0                                                                 &0 \cite{etaPVp}& \\
 &$K^{*0}\p^+$ &$-0.15\pm0.07$             &0                           &$0.004^{+0.013+0.043}_{-0.016-0.039}$   &$-0.01^{+0.01}_{-0.00}$ \cite{PVp}                                                                           &0\\
 &$K^{*+}\p^0$ &$-0.52\pm0.15$              &$-0.116\pm0.092$ &$0.016^{+0.031+0.111}_{-0.017-0.144}$    &$-0.32^{+0.21}_{-0.28}$ \cite{PVp}                                                                          &$-0.129^{+0.120}_{-0.122}\pm0.008$\\
 &$\r^0K^+$      &$0.37\pm0.11$               &$0.306\pm0.100$ &$0.454^{+0.178+0.314}_{-0.194-0.232}$   &$0.71^{+0.25}_{-0.35}$ \cite{PVp}                                                                            &$0.160^{+0.205+0.013}_{-0.224-0.016}$\\
 &$\r^+K^0$      &$0.21\pm0.36$             &0                           &$0.003^{+0.002+0.005}_{-0.003-0.002}$   &$0.01\pm0.01$ \cite{PVp}                                                                                        &0\\
 &$K^{*+}\e$    &$0.02\pm0.06$               &$-0.016\pm0.037$&$-0.101^{+0.039+0.065}_{-0.037-0.078}$ \cite{etaQ}  &$-0.2457^{+0.0072}_{-0.0027}$ \cite{etaKp}                                                           &$-0.019^{+0.034}_{-0.036}\pm0.001$\\
 &$K^{*+}\e'$     &$-0.26\pm0.27$             &$-0.391\pm0.162$&$0.697^{+0.065+0.279}_{-0.386-0.495}$ \cite{etaQ}   &$0.0460^{+0.0116}_{-0.0132}$ \cite{etaKp}                                                             &$0.026^{+0.267}_{-0.329}\pm0.002$\\
 &$\w K^+$        &$-0.02\pm0.04$             &$0.010\pm0.080$&$0.221^{+0.137+0.140}_{-0.128-0.130}$   &$0.32^{+0.15}_{-0.17}$ \cite{PVp}                                                                           &$0.123^{+0.166+0.008}_{-0.173-0.011}$\\
 &$\f K^+$          &$0.04\pm0.02$              &0                           &$0.006\pm0.001\pm0.001$                        &$0.01^{+0.00}_{-0.01}$ \cite{PVp}                                                                           &0\\
 &$K^{*+}\p^-$   &$-0.23\pm0.06$             &$-0.217\pm0.048$&$-0.121\pm0.005^{+0.126}_{-0.160}$        &$-0.60^{+0.32}_{-0.19}$ \cite{PVp}                                                                          &$-0.122^{+0.114}_{-0.113}\pm0.008$\\
 &$K^{*0}\p^0$  &$-0.15\pm0.13$             &$-0.332\pm0.114$&$-0.108^{+0.018+0.091}_{-0.028-0.063}$  &$-0.11^{+0.07}_{-0.05}$ \cite{PVp}                                                                          &$0.054^{+0.048+0.004}_{-0.051-0.005}$\\
 &$\r^-K^+$       &$0.20\pm0.11$               &$0.134\pm0.053$ &$0.319^{+0.115+0.196}_{-0.110-0.127}$   &$0.64^{+0.24}_{-0.30}$ \cite{PVp}                                                                            &$0.096^{+0.130+0.007}_{-0.135-0.009}$\\
 &$\r^0K^0$      &$0.06\pm0.20$               &$0.069\pm0.053$ &$0.087\pm0.012^{+0.087}_{-0.068}$         &$0.07^{+0.08}_{-0.05}$ \cite{PVp}                                                                           &$-0.035\pm0.048^{+0.003}_{-0.002}$\\
 &$K^{*0}\e$    &$0.19\pm0.05$               &$0.099\pm0.028$ &$0.034\pm0.004^{+0.027}_{-0.024}$ \cite{etaQ}   &$0.00570\pm0.00011$ \cite{etaKp}                                                                           &$-0.007^{+0.012+0.001}_{-0.013-0.000}$\\
 &$K^{*0}\e'$    &$-0.07\pm0.18$             &$0.069\pm0.152$ &$0.088^{+0.088+0.308}_{-0.107-0.241}$ \cite{etaQ}   &$-0.0130\pm0.0008$ \cite{etaKp}                                                                            &$0.099^{+0.062}_{-0.043}\pm0.009$\\
 &$\w K^0$       &$0.04\pm0.14$               &$-0.053\pm0.055$&$-0.047^{+0.018+0.055}_{-0.016-0.058}$  &$-0.03^{+0.02}_{-0.04}$ \cite{PVp}                                                                         &$0.038^{+0.052}_{-0.054}\pm0.003$\\
 &$\f K^0$         &$-0.01\pm0.14$             &0                            &$0.009^{+0.002+0.002}_{-0.001-0.001}$  &$0.03^{+0.01}_{-0.02}$ \cite{PVp}                                                                           &0 \\
\hline\hline
\end{tabular}
\caption{Same as Table \ref{preglobalBr} but for CP asymmetries.}
\label{preglobalAcp}
\end{table}
\endgroup

\begingroup
\squeezetable
\begin{table}[t!]
\begin{tabular}{ccccccc}
\hline\hline
\multicolumn{2}{c}{Mode}  & Data    & This Work   & QCDF    & pQCD & SCET             \\
\hline
$B^{+,0}\to$  &$\r^-\p^+$&$0.05\pm0.08$ &$-0.024\pm0.065$ &&$\sim 0.06$ \cite{Li:2014haa}     &\\
 &$\r^+\p^-$        &$0.07\pm0.14$             &$-0.049\pm0.074$ &&$\sim -0.22$ \cite{Li:2014haa}   &\\
 &$\r^0\p^0$       &$-0.23\pm0.34$               &$-0.229\pm0.112$&$-0.24^{+0.15+0.20}_{-0.14-0.22}$    & $\sim -0.30$ \cite{Li:2014haa}       &$-0.19\pm0.14^{+0.10}_{-0.15}$\\
 &$\r^0\e$           &-                                      &$-0.628\pm0.196$&$0.51^{+0.08+0.19}_{-0.07-0.32}$           &$0.227\pm0.061^{+0.139+0.096+0.236}_{-0.218-0.125-0.265}$ \cite{etaPVp}       &$0.29^{+0.36+0.09}_{-0.44-0.15}$\\
 &$\r^0\e'$          &-                                      &$-0.714\pm0.252$&$0.80^{+0.04+0.24}_{-0.09-0.43}$            &$-0.490^{+0.019+0.160+0.018+0.186}_{-0.008-0.081-0.042-0.178}$ \cite{etaPVp}&$0.38^{+0.22+0.09}_{-1.24-0.14}$\\
 &$\w\p^0$         &-                                      &$-0.315\pm0.195$ &$0.78^{+0.14+0.20}_{-0.20-1.39}$   & $\sim -0.26$ \cite{Li:2014haa}          &$0.72^{+0.36+0.07}_{-1.54-0.11}$\\
 &$\w\e$            &-                                      &$-0.461\pm0.113$&$-0.16\pm0.13^{+0.17}_{-0.16}$                &$0.390^{+0.003+0.506+0.059+0.029}_{-0.002-0.662-0.033-0.019}$ \cite{etaPVp} &$-0.16^{+0.14+0.10}_{-0.15-0.15}$\\
 &$\w\e'$           &-                                      &$-0.624\pm0.120$&$-0.28^{+0.14+0.16}_{-0.13-0.13}$           &$0.770^{+0.004+0.220+0.009+0.003}_{-0.001-0.529-0.001-0.000}$ \cite{etaPVp} &$-0.27^{+0.17+0.09}_{-0.33-0.14}$\\
 &$\f\p^0$         &-                                      &0                           & &&\\
 &$\f\e$             &-                                      &0                           &   &&\\
 &$\f\e'$            &-                                      &0                           &   &&\\
 &$\r^0K^0$      &$0.54^{+0.18}_{-0.21}$  &$0.643\pm0.036$ &$0.50^{+0.07+0.06}_{-0.14-0.12}$             &$0.50^{+0.10}_{-0.06}$ \cite{PVp}                                                                           &$0.56^{+0.02}_{-0.03}\pm0.01$\\
 &$\w K^0$       &$0.71\pm0.21$               &$0.789\pm0.028$ &$0.84\pm0.05^{+0.04}_{-0.06}$                 &$0.84^{+0.03}_{-0.07}$ \cite{PVp}                                                                           &$0.80\pm0.02\pm0.01$\\
 &$\f K^0$         &$0.74^{+0.11}_{-0.13}$  &$0.718\pm0.000$ &$0.692^{+0.003}_{-0.000}\pm0.002$&$0.71\pm0.01$ \cite{PVp}                                                                         &0.69\\
\hline\hline
\end{tabular}
\caption{Same as Table \ref{preglobalBr} but for the time-dependent CP asymmetry $\cal S$.}
\label{preglobalS}
\end{table}
\endgroup

With reference to Table~\ref{preglobalAcp}, all the theoretical calculations  predict a negative CP asymmetry for the $B^+\to\rho^0\pi^+$ decay, whereas the current data has a positive central value.  It is thus interesting to see what future data will be when the uncertainties are reduced.  It should be stressed that BaBar has found the first evidence of direct CP violation in the decay $B^+\to K^{*+}\pi^0$, $A_{CP}=-0.52\pm0.14\pm0.04\pm0.04$, from the preliminary analysis of $B^+\to K_S\pi^+\pi^0$ decay \cite{etaKexp}.  Our prediction of $A_{CP}=-0.116\pm0.092$ is substantially smaller. Hence, it is important to have independent measurements of CP asymmetry for this mode. Note that the predicted $A_{CP}(K^{*+}\pi^0)$ by QCDF \cite{Q} has a wrong sign when confronted with experiment. As discussed before, this may be attributed to the large complex $P_{EW,V}$ amplitude whose effect was not considered in \cite{Q}.
Theories have diverse predictions in the sign and/or magnitude of several asymmetries, such as the $B^+ \to \rho^+ \eta$ and $K^{*+} \eta'$ modes and $B^0 \to K^{*0} \eta$ mode.  Therefore, a better experimental determination of these observables will be very useful in checking theory calculations.  There also exist diverse predictions for the CP asymmetries of the $\rho^0 \eta^{(\prime)}$, $\omega \pi^0$, $\omega \eta^{(\prime)}$ modes.  Yet a measurement of them in the near future is unlikely due to their small branching fractions.  Predictions for the time-dependent CP asymmetries $\cal S$ of all the $B^{+,0}$ decays are given in Table~\ref{preglobalS}, where one also observes diverse predictions for the more difficult $\rho^0 \eta^{(\prime)}$, $\omega \pi^0$, $\omega \eta^{(\prime)}$ modes.

\begingroup
\begin{table}[t!]
\begin{tabular}{cccccc}
\hline\hline
\multicolumn{2}{c}{Mode}  & This Work    & QCD     & pQCD     &SCET         \\
\hline
$B_s\to$&$\o K^{*0}\p^0$&$3.07\pm1.20$&$0.89^{+0.80+0.84}_{-0.34-0.35}$&$0.07^{+0.02+0.04}_{-0.01-0.02}\pm0.01$      &$1.07^{+0.16+0.10}_{-0.15-0.09}$  \\
 &$K^{*-}\p^+$     &$7.92\pm1.02$       &$7.8^{+0.4+0.5}_{-0.7-0.7}$            &$7.6^{+2.9+0.4+0.5}_{-2.2-0.5-0.3}$                &$6.6^{+0.2}_{-0.1}\pm0.7$  \\
 &$\r^+K^-$         &$14.63\pm1.46$      &$14.7^{+1.4+0.9}_{-1.9-1.3}$          &$17.8^{+7.7+1.3+1.1}_{-5.6-1.6-0.9}$              &$10.2^{+0.4}_{-0.5}\pm0.9$  \\
 &$\r^0\o K^0$    &$0.56\pm0.24$        &$1.9^{+2.9+1.4}_{-0.9-0.6}$            &$0.08\pm0.02^{+0.07+0.01}_{-0.03-0.00}$       &$0.81^{+0.05+0.08}_{-0.02-0.09}$  \\
 &$\o K^{*0}\e$   &$1.44\pm0.54$        &$0.56^{+0.33+0.35}_{-0.14-0.17}$  &$0.17\pm0.04^{+0.10+0.03}_{-0.06-0.01}$       &$0.62\pm0.14^{+0.07}_{-0.08}$  \\
 &$\o K^{*0}\e'$  &$1.65\pm0.60$        &$0.90^{+0.69+0.72}_{-0.30-0.41}$   &$0.09\pm0.02^{+0.03}_{-0.02}\pm0.01$           &$0.87^{+0.35+0.10}_{-0.32-0.08}$  \\
 &$\w\o K^0$      &$0.58\pm0.25$        &$1.6^{+2.2+1.0}_{-0.7-0.5}$             &$0.15^{+0.05+0.07+0.02}_{-0.04-0.03-0.01}$  &$1.3\pm0.1\pm0.1$  \\
 &$\f\o K^0$        &$0.41\pm0.07$        &$0.6^{+0.5+0.4}_{-0.2-0.3}$            &$0.16^{+0.04+0.09+0.02}_{-0.03-0.04-0.01}$   &$0.54^{+0.21+0.08}_{-0.17-0.07}$  \\
 &$K^{*+}K^-$     &$8.03\pm0.48$        &$10.3^{+3.0+4.8}_{-2.2-4.2}$          &$6.0^{+1.7+1.7+0.7}_{-1.5-1.2-0.3}$                 &$9.5^{+3.2+1.2}_{-2.8-1.1}$  \\
 &$K^{*-}K^+$     &$7.98\pm0.77$        &$11.3^{+7.0+8.1}_{-3.5-5.1}$          &$4.7^{+1.1+2.5}_{-0.8-1.4}\pm0.0$                    &$10.2^{+3.8+1.5}_{-3.2-1.2}$  \\
 &$K^{*0}\o K^0$ &$9.33\pm0.54$       &$10.5^{+3.4+5.1}_{-2.8-4.5}$          &$7.3^{+2.5+2.1}_{-1.7-1.3}\pm0.0$                    &$9.3^{+3.2+1.2}_{-2.8-1.0}$  \\
 &$\o K^{*0}K^0$ &$6.32\pm0.68$       &$10.1^{+7.5+7.7}_{-3.6-4.8}$          &$4.3\pm0.7^{+2.2}_{-1.4}\pm0.0$                      &$9.4^{+3.7+1.4}_{-3.1-1.2}$  \\
 &$\r^0\e$            &$0.34\pm0.21$       &$0.10^{+0.02+0.02}_{-0.01-0.01}$  &$0.06^{+0.03}_{-0.02}\pm0.01\pm0.00$            &$0.06^{+0.03}_{-0.02}\pm0.00$  \\
 &$\r^0\e'$           &$0.31\pm0.19$       &$0.16^{+0.06}_{-0.02}\pm0.03$      &$0.13^{+0.06}_{-0.04}\pm0.02^{+0.00}_{-0.01}$&$0.14^{+0.24}_{-0.11}\pm0.01$  \\
 &$\w\e$              &$0.15\pm0.16$       &$0.03^{+0.12+0.06}_{-0.02-0.01}$  &$0.04^{+0.03+0.05}_{-0.01-0.02}\pm0.00$        &$0.007^{+0.011}_{-0.002}\pm0.001$  \\
 &$\w\e'$             &$0.14\pm0.14$       &$0.15^{+0.27+0.15}_{-0.08-0.06}$  &$0.44^{+0.18+0.15+0.00}_{-0.13-0.14-0.01}$    &$0.20^{+0.34}_{-0.17}\pm0.02$  \\
 &$\f\p^0$            &$1.94\pm1.14$      &$0.12^{+0.02+0.04}_{-0.01-0.02}$   &$0.16^{+0.06}_{-0.05}\pm0.02\pm0.00$            &$0.09\pm0.00\pm0.01$  \\
 &$\f\e$                &$0.39\pm0.39$      &$1.0^{+1.3+3.0}_{-0.1-1.2}$             &$3.6^{+1.5+0.8}_{-1.0-0.6}\pm0.0$                    &$0.94^{+1.89+0.16}_{-0.97-0.13}$  \\
 &$\f\e'$               &$5.48\pm1.84$      &$2.2^{+4.5+8.3}_{-1.9-2.5}$             &$0.19^{+0.06+0.19}_{-0.01-0.13}\pm0.00$        &$4.3^{+5.2+0.7}_{-3.6-0.6}$  \\
\hline\hline
\end{tabular}
\caption{Predicted branching fractions in units of $10^{-6}$ for all the $B_s$ decays using the fit results of Scheme B. Predictions made by QCDF, pQCD and SCET are obtained from Ref.~\cite{sQ}, Ref.~\cite{sp} and Ref.~\cite{PVS} (work 2), respectively. }
\label{preglobalsBr}
\end{table}
\endgroup

We next turn to the $B_s$ sector.
Although evidence of the $B_s\to K^{*-}\pi^+$ and $K^{*\pm}K^\mp$ decays have been seen, none of the $B_s \to VP$ decays have been firmly established yet.  In Table~\ref{preglobalsBr}, theoretical calculations differ in the branching fractions of the $\o K^{*0} \pi^0$, $\rho^+ K^-$, $\o K^{*0} \eta$, $\o K^{*0} \eta'$, $\omega \o K^0$, $\phi \o K^0$, $\phi \pi^0$ and $\phi \eta'$ modes.  In particular, the predicted branching fractions of $\o K^{*0}\pi^0$ and $\phi\pi^0$ modes in our work are much larger than those made by the theoretical calculations based on the short-distance effective Hamiltonian.  This is mainly because we have a large $C_V$ amplitude involved in both modes and a large $P_{EW,V}$ amplitude for the latter.
The decay $B_s\to \phi\pi^0$ is governed by the $c'_V$ amplitude. As explained in Section V-A, since $|Y_{sb}^c| \gg |Y_{sb}^u|$ and $|P_{EW,V}| \sim |P_V|$, the $P_{EW,V}$ amplitude makes a substantial contribution to $c'_V$ and enhances $BF(B_s\to \phi\pi^0)$ to the level of $2\times 10^{-6}$. It has been claimed in the literature \cite{Hofer} that if this decay mode is observed at the level of $10^{-6}$, it will be a signal of new physics effect on $P_{EW,V}$. In our diagrammatic analysis of the experimental data, $P_{EW,V}$ is found to be large and complex. Whether or not this is related to new physics is another issue which will not be addressed here.

\begingroup
\begin{table}[t!p]
\begin{tabular}{cccccc}
\hline\hline
\multicolumn{2}{c}{Mode}  & This Work      & QCDF    &pQCD   &SCET             \\
\hline
$B_s\to$&$\o K^{*0}\p^0$&$-0.423\pm0.158$ &$-0.263^{+0.108+0.422}_{-0.109-0.367}$&$-0.471^{+0.074+0.355+0.029}_{-0.087-0.298-0.070}$&$0.134^{+0.186+0.008}_{-0.188-0.012}$ \\
 &$K^{*-}\p^+$      &$-0.136\pm0.053$ &$-0.240^{+0.012+0.077}_{-0.015-0.039}$       &$-0.190^{+0.025+0.027+0.009}_{-0.026-0.034-0.014}$ &$-0.124^{+0.175+0.011}_{-0.153-0.012}$ \\
 &$\r^+K^-$           &$0.120\pm0.027$  &$0.117^{+0.035+0.101}_{-0.021-0.116}$         &$0.142^{+0.024+0.023+0.012}_{-0.022-0.016-0.007}$  &$0.108^{+0.094+0.009}_{-0.102-0.010}$ \\
 &$\r^0\o K^0$      &$-0.124\pm0.453$ &$0.289^{+0.146+0.250}_{-0.145-0.237}$        &$0.734^{+0.064+0.162+0.022}_{-0.117-0.478-0.039}$   &$-0.325^{+0.307+0.027}_{-0.234-0.029}$ \\
     &                          &$-0.348\pm0.285$ &$0.29^{+0.23+0.16}_{-0.24-0.21}$                  &$-0.57^{+0.22+0.51+0.02}_{-0.17-0.39-0.05}$                &$-0.03^{+0.22+0.17}_{-0.17-0.12}$ \\
 &$\o K^{*0}\e$     &$0.828\pm0.123$  &$0.400^{+0.111+0.531}_{-0.192-0.645}$         &$0.512^{+0.062+0.141+0.020}_{-0.064-0.124-0.033}$   &$-0.627^{+0.281+0.026}_{-0.225-0.039}$ \\
 &$\o K^{*0}\e'$    &$-0.408\pm0.273$ &$-0.625^{+0.060+0.247}_{-0.055-0.202}$       &$-0.511^{+0.046+0.150+0.032}_{-0.066-0.182-0.041}$   &$-0.321^{+0.228+0.026}_{-0.232-0.017}$ \\
 &$\w\o K^0$         &$-0.029\pm0.436$&$-0.320^{+0.189+0.236}_{-0.175-0.262}$        &$-0.521^{+0.032+0.227+0.032}_{-0.000-0.151-0.020}$   &$0.182^{+0.164+0.012}_{-0.170-0.017}$ \\
     &                         &$0.928\pm0.110$  &$0.92^{+0.03+0.08}_{-0.07-0.15}$&$-0.63\pm0.09^{+0.28+0.01}_{-0.11-0.02}$                  &$0.98^{+0.02+0.00}_{-0.04-0.01}$ \\
 &$\f\o K^0$         &0                            &$-0.032^{+0.012+0.006}_{-0.014-0.013}$        &0                                                                                        &$-0.022^{+0.030}_{-0.029}\pm0.001$ \\
     &                         &$-0.692\pm0.000$ &$-0.69\pm0.01\pm0.01$         & $-0.72$                                                          &$-0.13\pm0.02\pm0.01$ \\
 &$K^{*+}K^-$      &$-0.217\pm0.048$&$-0.110^{+0.005+0.140}_{-0.004-0.188}$         &$-0.366\pm0.023^{+0.028+0.013}_{-0.035-0.012}$         &$-0.123^{+0.114}_{-0.113}\pm0.008$ \\
 &$K^{*-}K^+$      &$0.134\pm0.053$ &$0.255^{+0.092+0.163}_{-0.088-0.113}$          &$0.553^{+0.044+0.085+0.051}_{-0.049-0.098-0.025}$    &$0.096^{+0.130+0.007}_{-0.135-0.009}$ \\
 &$K^{*0}\o K^0$ &0                           &$0.0049^{+0.0008+0.0009}_{-0.0007-0.0012}$&0                                                                                        &0 \\
 &$\o K^{*0}K^0$ &0                           &$0.0010^{+0.0008+0.0005}_{-0.0007-0.0002}$&0                                                                                        &0 \\
 &$\r^0\e$            &$0.323\pm0.136$ &$0.757^{+0.153+0.133}_{-0.176-0.375}$          &$-0.092^{+0.010+0.028+0.004}_{-0.004-0.027-0.007}$   &0 \\
     &                         &$-0.002\pm0.168$&$0.35^{+0.09+0.22}_{-0.16-0.40}$                    &$0.15\pm0.06^{+0.14}_{-0.16}\pm0.01$                           &$0.60^{+0.30}_{-0.53}\pm0.03$ \\
 &$\r^0\e'$           &$0.323\pm0.136$ &$0.874^{+0.034+0.057}_{-0.106-0.303}$           &$0.258^{+0.013+0.028+0.034}_{-0.020-0.036-0.015}$    &0 \\
     &                        &$-0.002\pm0.168$  &$0.45^{+0.05+0.30}_{-0.13-0.35}$                     &$-0.16\pm0.00^{+0.10+0.04}_{-0.12-0.05}$                     &$-0.41\pm0.75^{+0.10}_{-0.15}$ \\
 &$\w\e$              &$-0.432\pm0.271$&$-0.648^{+0.244+0.440}_{-0.034-0.316}$          &$-0.167^{+0.058+0.154+0.008}_{-0.032-0.191-0.017}$   &0 \\
     &                        &$-0.238\pm0.296$&$-0.76^{+0.16+0.52}_{-0.03-0.22}$                     &$-0.02^{+0.01+0.02}_{-0.03-0.08}\pm0.00$                     &$0.93^{+0.04+0.03}_{-0.98-0.04}$ \\
 &$\w\e'$             &$-0.432\pm0.271$&$-0.394^{+0.044+0.104}_{-0.030-0.117}$           &$0.077^{+0.004+0.045+0.094}_{-0.001-0.042-0.004}$    &0 \\
     &                        &$-0.238\pm0.296$&$-0.84^{+0.06+0.04}_{-0.05-0.03}$                     &$-0.11^{+0.01}_{-0.00}\pm0.04^{+0.02}_{-0.03}$             &$-1.00^{+0.04+0.01}_{-0.00-0.00}$ \\
 &$\f\p^0$           &$0.073\pm0.201$ &$0.822^{+0.109+0.090}_{-0.140-0.553}$             &$0.133^{+0.003+0.021+0.015}_{-0.004-0.017-0.007}$    &0 \\
     &                        &$0.439\pm0.171$ &$0.40^{+0.04+0.32}_{-0.10-0.53}$                       &$-0.07\pm0.01^{+0.08+0.02}_{-0.09-0.03}$                     &$0.90\pm0.00^{+0.02}_{-0.03}$ \\
 &$\f\e$               &$0.428\pm0.504$ &$-0.124^{+0.141+0.649}_{-0.057-0.398}$            &$-0.018^{+0.000}_{-0.001}\pm0.006^{+0.001}_{-0.002}$&$0.169^{+0.138}_{-0.183}\pm0.016$ \\
     &                        &$0.534\pm0.400$&$0.21^{+0.08+0.61}_{-0.11-0.25}$                       &$-0.03^{+0.02+0.07+0.01}_{-0.01-0.20-0.02}$                 &$0.23^{+0.35}_{-0.16}\pm0.02$ \\
 &$\f\e'$              &$0.043\pm0.090$ &$0.139^{+0.154+0.285}_{-0.042-0.897}$             &$0.078^{+0.015+0.012+0.001}_{-0.005-0.086-0.004}$    &$0.078^{+0.050}_{-0.049}\pm0.008$ \\
     &                        &$0.166\pm0.057$ &$0.08^{+0.05+0.48}_{-0.06-0.81}$                       &$0.00\pm0.00\pm0.02\pm0.00$                                       &$0.10^{+0.07}_{-0.05}\pm0.01$ \\
\hline\hline
\end{tabular}
\caption{Same as Table~\ref{preglobalsBr} but for CP asymmetries.  Whenever there exists more than one line, the upper line is $\cal A$ while the second line is $\cal S$.}
\label{preglobalsAcp}
\end{table}
\endgroup

All the theory predictions on $BF(B_s \to K^{*-} \pi^+)$ are consistent with one another, but more than twice larger than the central value of current data.  In fact, the $B_s \to K^{*-} \pi^+$ decay and the $B^0 \to \rho^- \pi^+$ should have the same decay width when the $W$-exchange amplitude is ignored.  With roughly the same lifetime for the two neutral $B$ mesons, we therefore expect $BF(B_s \to K^{*-} \pi^+) \simeq BF(B^0 \to \rho^- \pi^+) \simeq 8 \times 10^{-6}$  and $A_{CP}(B_s \to K^{*-} \pi^+) \simeq {\cal A}(B^0 \to \rho^- \pi^+) \simeq 0.14$.  Likewise,  $BF(B_s \to \rho^+ K^-) \simeq BF(B^0 \to \rho^+ \pi^-) \simeq 15 \times 10^{-6}$ and $A_{CP}(B_s \to \rho^+ K^-) \simeq {\cal A}(B^0 \to \rho^+ \pi^-) \simeq 0.12$.  Similar patterns also exist in the $|\Delta S| = 1$ transitions for the following two sets of modes: $B^0 \to K^{*+} \pi^-$ and $B_s \to K^{*+} K^-$; and $B^0 \to \rho^- K^{+}$ and $B_s \to K^{*-} K^+$.  It is also noted that all predictions and the measured value of $BF(B_s \to K^{*\pm} K^\mp)$ are in good agreement within errors.

Turning to CP asymmetries in Table~\ref{preglobalsAcp}, theories have diverse predictions for $\cal A$ and $\cal S$ of $B_s\to \rho^0 \o K^0$ and $\rho^0 \eta$.  Most predict ${\cal S}(\omega \o K^0)$ close to 1, yet pQCD has an opposite sign at $\sim -0.6$.  Most predict ${\cal S}(\phi \o K^0) \simeq -0.7$, whereas SCET predicts it to be $\sim -0.1$.  Both our work and QCDF predict ${\cal S}(\phi \pi^0) \simeq 0.4$, significantly different from those of pQCD and SCET.  However, these observables are difficult to measure because of the small branching fractions except for possibly the $\phi \pi^0$ mode.

\section{Comparison with Factorization for $a_{1,2}$ \label{sec:PPQCDFsec}}

\begingroup
\begin{table}[t]
\begin{tabular}{ccccccccc}
\hline\hline
   & ~~~~$\pi^+\pi^0$~~~~ & ~~~~$K^+\pi^0$~~~~ & ~~~~$\rho^0\pi^+$~~~~ & ~~~~$\rho^+\pi^0$~~~~
   & ~~~~$\w\pi^+$~~~~ & ~~~~$K^{*+}\pi^0$~~~~ & ~~~~$\rho^0 K^+$~~~~ & ~~~~$\w K^+$~~~~ \\
\hline
 $|a_1|$       &$0.82\pm0.02$ &$0.67\pm0.01$&$0.97\pm0.07$&$0.96\pm0.05$&$1.05\pm0.07$&$0.95\pm0.05$&$0.80\pm0.05$&$0.86\pm0.06$\\
 $|a_2|$       &$0.66\pm0.06$ &$0.47\pm0.05$&$0.28\pm0.11$&$0.74\pm0.47$&$0.32\pm0.13$&$0.60\pm0.27$&$0.20\pm0.08$&$0.23\pm0.09$\\
 $|a_2/a_1|$&$0.80\pm0.08$ &$0.70\pm0.07$&$0.29\pm0.11$&$0.77\pm0.35$&$0.31\pm0.12$&$0.63\pm0.29$&$0.25\pm0.10$&$0.27\pm0.11$\\
\hline\hline
\end{tabular}
\caption{The extracted parameters $a_1$ and $a_2$ from $B^+$ decays in Scheme A of both the $PP$ and $VP$ sectors.}
\label{a1a2}
\end{table}
\endgroup

In the factorization approach, the color-allowed tree amplitude and the color-suppressed tree amplitude for a $B \to M_1 M_2$ decay can be computed as follows:
\begin{equation}
 \begin{split}
 T_{M_1} &=\frac{G_F}{\sqrt2} a_1(M_1M_2) X^{(BM_1,M_2)} ~,\\
 C_{M_1} &=\frac{G_F}{\sqrt2} a_2(M_1M_2) X^{(BM_1,M_2)} ~,\\
 \end{split}
 \label{form}
 \end{equation}
where
$M_1$ and $M_2$ can be pseudoscalar or vector mesons, and $a_1(M_1M_2)$ and $a_2(M_1M_2)$ are the effective Wilson coefficients.  The hadronic matrix element $X^{(\o BM_1,M_2)}$, denoted by $A_{M_1M_2}$ in \cite{a1a2pipi}, can be factorized into a product of decay constant and form factor:
 \begin{equation}
 \begin{split}
X^{(BP_1,P_2)}&=if_{P_2}(m_B^2-m_{P_1}^2)F_0^{BP_1}(m_{P_2}^2) ~,\\
X^{(BP,V)}&=2f_Vm_Bp_cF_1^{BP}(m_V^2) ~,\\
X^{(BV,P)}&=2f_Pm_Bp_cA_0^{BV}(m_P^2) ~.\\
 \end{split}
 \label{X}
 \end{equation}
For numerical calculations, we take the decay constants and form factors given in Ref.~\cite{QCDFinputs}.

In Table~\ref{a1a2}, we list $|a_{1,2}|$ and the ratio $|a_2/a_1|$ for various modes as extracted based on Scheme A for both $PP$ and $VP$ modes in our analyses.  Because of SU(3) breaking effects in meson masses, decay constants and form factors, the extracted parameters $a_{1,2}$ vary from channel to channel.
This has the advantage of a more direct comparison with the effective Wilson coefficients calculated in the perturbative approach.  For example, perturbative calculations have $|a_1(\pi^+\pi^0)|=1.015\pm0.024$ and $|a_2(\pi^+\pi^0)|=0.218\pm0.103$~\cite{Bell}, while our work has $|a_1(\pi^+\pi^0)|$ and $|a_2(\pi^+\pi^0)|$ to be $0.82\pm0.02$ and $0.66\pm0.06$, respectively.  From the table, it is seen that $|a_2/a_1|$ is larger than $\sim 0.7$ in the $PP$ sector.  In the $VP$ sector, the $|a_2/a_1|$ is around $0.3$ for decay modes involving $T_V$ and $C_P$, yet larger than $\sim 0.7$ for those involving $T_P$ and $C_V$.

\section{Conclusions \label{sec:conclusions}}

In order to make predictions for all of the $B\to PP$ and $B\to PV$ decays, a set of theory parameters has to be determined from experiment.  To achieve this goal, we have performed $\chi^2$ fitting within the framework of the diagrammatic approach based on flavor SU(3) symmetry.  We have obtained the 1-$\sigma$ ranges of each theory parameter and used them to make predictions.

The main results of the present work are:

\begin{itemize}

\item
In the $PP$ sector, the color-suppressed tree amplitude $C$ is found to be larger than previously known and has a strong phase of $\sim -70^\circ$ relative to the color-favored tree amplitude $T$.
We have extracted for the first time the $W$-exchange $E$ and penguin-annihilation $P\!A$ amplitudes.  The former has a size of about the QCD-penguin amplitude and a phase opposite to that of $T$, while the latter is suppressed in magnitude but gives the dominant contribution to the $B_s^0\to \pi^+\pi^-$ and $\pi^0\pi^0$ decays due to the enhancement in CKM matrix elements.

\item
The flavor-singlet amplitude for decays involving SU(3)$_F$-singlet mesons plays an essential role particularly in explaining the branching fractions of the $\eta' K$ decays.   The associated phase is $\sim -100^\circ$ with respect to the $T$ amplitude.  The branching fraction of $B^0 \to \eta'\pi^0$ is predicted much larger than other theory predictions and closer to the measured value due to a constructive interference between the QCD penguin and flavor-singlet diagrams, which subtend a phase less than $90^\circ$.

\item
The ratio $|C/T|$ has values $\agt 0.7$. It is tempting to conjecture that such a large $|C|$ could be attributed to some particular set of observables, such as the $B^0 \to \pi^0\pi^0$ and/or $K^0 \pi^0$ decays.  We have examined this issue and  found that the large $|C|$ is required not just by any individual modes mentioned above.  We have shown that a large complex $C$ results from a fit to the observed direct CP violation  in $B\to K\pi$ decays.

\item
We have tested flavor SU(3) symmetry, a working principle in the present work, by allowing symmetry breaking factors in the decay amplitudes, and found that it is indeed a good approximate symmetry.

\item
In the $VP$ sector, the color-suppressed tree amplitude $C_V$  with the spectator quark ending up in the vector meson has a large size and a strong phase of $\sim -90^\circ$ relative to the color-favored tree amplitudes.  The associated electroweak penguin amplitude $P_{EW,V}$ also has a similar strong phase and a magnitude comparable to the corresponding QCD penguin amplitude $P_V$.  In contrast, the color-suppressed tree, QCD penguin, and electroweak penguin amplitudes with the spectator quark ending up in the pseudoscalar meson have magnitudes more consistent with na{\"i}ve expectations.  Besides, current data are not sufficiently precise for us to fix the $W$-exchange amplitudes.

\item
The observation of the $P_{EW,V}$ and $P_V$ amplitudes comparable in magnitude has some important implications. For example, it explains why the CP asymmetries of $B^+\to K^{*+}\pi^0$ and $B^0\to K^{*+}\pi^-$ are of the same sign and predicts a large branching fraction of $B_s\to\phi\pi^0$ at about $2\times 10^{-6}$, one order of magnitude larger than conventional theory predictions.

\item
For both the $PP$ and $VP$ sectors, predictions of all the decay modes are made based upon our fit results and compared with data and those made by perturbative approaches.  We have identified a few observables to be determined experimentally in order to discriminate among theory calculations.

\end{itemize}

\section*{Acknowledgments}

C.-W. C would like to thank the hospitality of KITP at Santa Barbara, where this work was finalized.  This research was supported in part by the Ministry of Science and Technology of R.O.C. under Grant Nos. 100-2112-M-001-009-MY3 and 100-2628-M-008-003-MY4, and in part by the National Science Foundation under Grant No. NSF PHY11-25915.


\end{document}